\documentclass[aps,twocolumn,superscriptaddress]{revtex4-1}

\usepackage{amsmath,amssymb,color,graphicx,mathpazo,mathtools,times,float,dsfont}
\usepackage[colorlinks=true,allcolors=blue]{hyperref}
\usepackage{soul}
\usepackage{comment}
\definecolor{Red}{rgb}{1,0,0}

\usepackage[textsize=scriptsize]{todonotes}

\newcommand{\eref}[1]{Eq.~\eqref{eq:#1}}

\newcommand{\Fref}[1]{Figure~\ref{fig:#1}}
\newcommand{\fref}[1]{Fig.~\ref{fig:#1}}

\begin{document}
\title{Floquet phonon lasing in multimode optomechanical systems}

\author{Laura Mercad\'e}\thanks{These authors contributed equally to this work}
\affiliation{Nanophotonics Technology Center, Universitat Polit\`ecnica de Valencia, Camino de Vera s/n, 46022 Valencia, Spain}
\author{Karl Pelka}\thanks{These authors contributed equally to this work}
\affiliation{Department of Physics, University of Malta, Msida MSD 2080, Malta}
\author{Roel Burgwal}
\affiliation{Department of Applied Physics and Institute of Photonic Integration, Eindhoven University of Technology, P.O. Box 513, 5600 MB Eindhoven, The Netherlands}
\affiliation{Center for Nanophotonics, AMOLF, Science Park 104, 1098 XG Amsterdam, The Netherlands}
\author{Andr\'e Xuereb}
\affiliation{Department of Physics, University of Malta, Msida MSD 2080, Malta}
\author{Alejandro Mart\'inez}
\affiliation{Nanophotonics Technology Center, Universitat Polit\`ecnica de Valencia, Camino de Vera s/n, 46022 Valencia, Spain}
\author{Ewold Verhagen}
\email{verhagen@amolf.nl}
\affiliation{Center for Nanophotonics, AMOLF, Science Park 104, 1098 XG Amsterdam, The Netherlands}
\affiliation{Department of Applied Physics and Institute of Photonic Integration, Eindhoven University of Technology, P.O. Box 513, 5600 MB Eindhoven, The Netherlands}

\date{\today}

\begin{abstract}
Dynamical radiation pressure effects in cavity optomechanical systems give rise to self-sustained oscillations or `phonon lasing' behavior, producing stable oscillators up to GHz frequencies in nanoscale devices. Like in photonic lasers, phonon lasing normally occurs in a single mechanical mode. We show here that phase-locked, multimode phonon lasing can be established in a multimode optomechanical system through Floquet dynamics induced by a temporally modulated laser drive. We demonstrate this concept in a suitably engineered silicon photonic nanocavity coupled to multiple GHz-frequency mechanical modes. We find that the long-term frequency stability is significantly improved in the multimode lasing state as a result of the phase locking. These results provide a path towards highly stable ultra-compact oscillators, pulsed phonon lasing, coherent waveform synthesis, and emergent many-mode phenomena in oscillator arrays.
\end{abstract}

\maketitle

\emph{Introduction.}---Recent times have seen extraordinary progress in exploiting radiation pressure control over optical and mechanical degrees of freedom in cavity optomechanical systems~\cite{Aspelmeyer2014}. The combined advantages of high mechanical coherence and quantum-noise-limited optical control allow the generation of pure quantum states of macroscopic mechanical resonators~\cite{Chan2011,Teufel2011} and their use as quantum transducers~\cite{RE11-JPC}. The very same advantages lead to highly coherent self-oscillations associated with a parametric instability~\cite{Kippenberg2005, KAR04-NAT}. Above threshold, a blue-detuned optical drive induces phonon lasing that can be used for timekeeping, microwave oscillators, signal synthesis, narrowband filters, and studying nonlinear dynamics~\cite{Marquardt2006, Eichenfield2009, NAV14-AIP, Mercade2020, GHOR19-APLPhot}. 

Even if multiple mechanical modes are coupled to a cavity, phonon lasing takes place for a single mode whose threshold condition is satisfied first, whilst the other modes get cooled~\cite{Kemiktarak2014} --- similar to gain suppression (mode competition) in regular lasers~\cite{LAMB64-PR}. In optical lasers and parametric oscillators, mode-locking is achieved through temporal control techniques such as synchronous pumping \cite{HALL93-JOS}. Simultaneous phononic self-oscillation in multiple modes was observed in a low-Finesse cavity, but without phase locking~\cite{MET08-PRL}. A reliable route to phase-locked phonon lasing opens the door to versatile optomechanical signal synthesis, which is especially valuable at high (GHz) frequencies in chip-scale devices. Moreover, it has broad significance in view of the emergent phenomena in multimode self-oscillating systems, including synchronization~\cite{Heinrich2011,Lipson2012,Holmes2012,Loerch2017,Colombano2019,Pelka2020,Madiot2020}, stability enhancement~\cite{Lipson2015}, dynamical topological phases~\cite{Walter2016}, and analog simulators~\cite{MAH16-SA}.

In this work, we show that the single-mode lasing limitation can be overcome by using a Floquet approach in which the optical drive is modulated in time. Recently, time-modulated radiation pressure was used to couple mechanical modes of different frequencies and enable mechanical state transfer~\cite{Weaver2017}, nonreciprocity~\cite{Xu2019}, synthetic gauge fields~\cite{Mathew2018}, and entanglement~\cite{Ockeloen2018}. Establishing a Floquet theory for phonon lasing, we show that a laser drive modulated at the difference between two mechanical modes' frequencies can induce phase-locked coherent oscillation of both. We observe the predicted multimode lasing experimentally in a silicon optomechanical crystal cavity supporting two GHz-frequency mechanical modes. We also find that the long-term stability of the output microwave tones is significantly improved, promising exploitation of these mechanisms towards highly stable, ultra-compact oscillators for microwave photonics. 

\begin{figure}
\begin{center}
\includegraphics[width=0.49\textwidth]{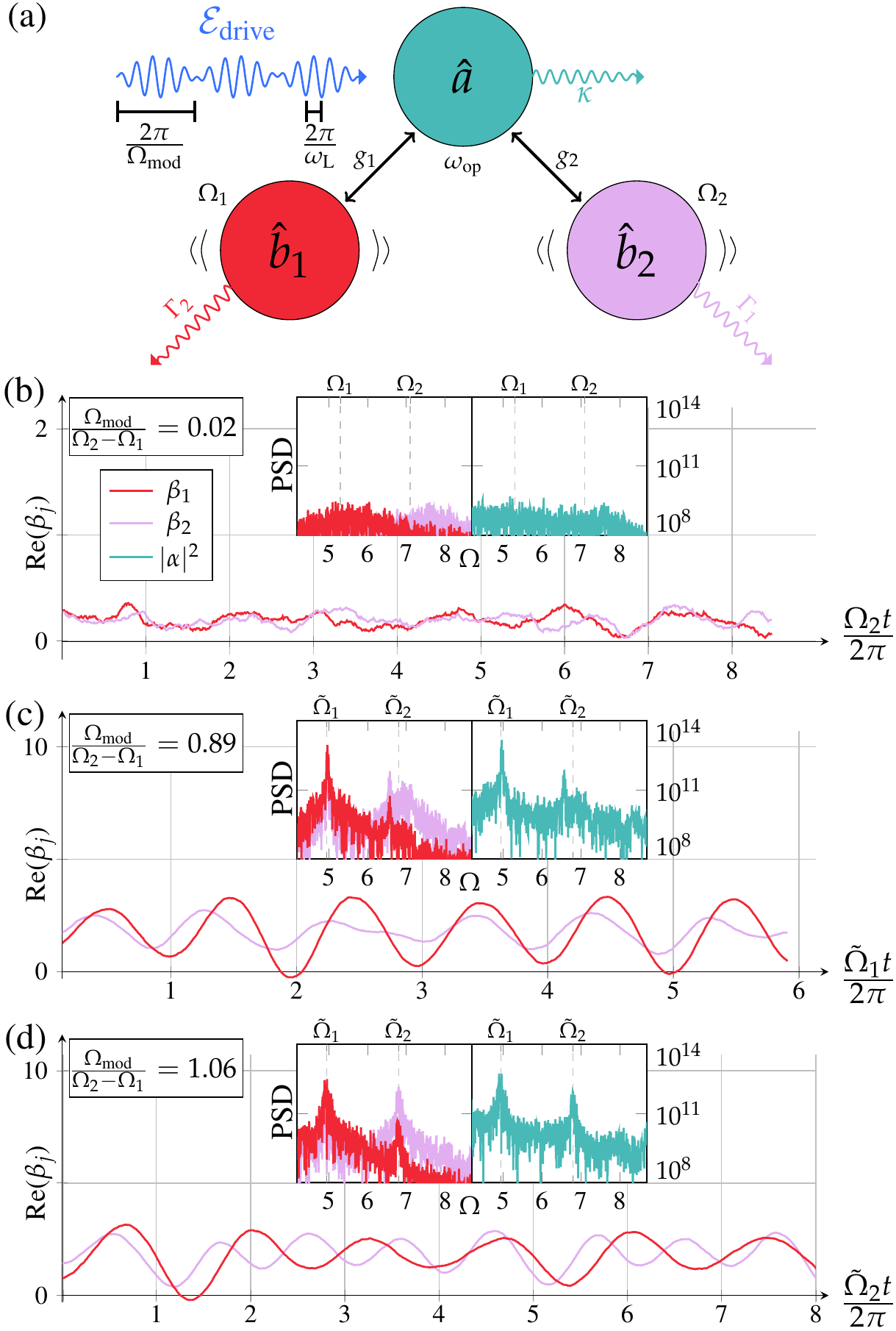}%
\end{center}
\caption{(Color online) Multimode phonon lasing in an optomechanical cavity. (a) Two mechanical modes at distinct frequencies are coupled via radiation pressure to an optical cavity driven by an intensity modulated pump. (b) For modulation frequency much smaller than the difference frequency both mechanical modes are in a thermal state. (c) With the modulation frequency approaching the difference frequency one mode starts self-sustained oscillations (SSO) with the help of the multimode gain mechanism. (d) When the modulation frequency is approximately the difference frequency, both mechanical modes coherently oscillate, i.e. multimode oscillation (MMO) occurs. Insets show corresponding power spectral densities of the mechanical (left) and optical (right) modes. Modulation frequencies are indicated in the panels, other parameters in the main text.}
\label{fig:Numerics1}
\end{figure}

\emph{Theoretical model.}---We consider the collective dynamics of a system consisting of $N$ mechanical modes (labeled by $j$) coupled to one optical mode, described by the Hamiltonian
\begin{align}
\label{eq:OMModel}
\hat{H}_{\text{S}}/\hbar=\omega_{\text{op}}\hat{a}^{\dagger}\hat{a}+\sum\limits_{j=1}^N\bigl[\Omega_j\hat{b}_j^{\dagger}\hat{b}_j-g_{j}\hat{a}^{\dagger}\hat{a}(\hat{b}_j+\hat{b}_j^{\dagger})\bigr],
\end{align}
with $\hat{a}$ ($\hat{b}_j$) the optical (mechanical) annihilation operator, $\omega_\text{op}$ ($\Omega_j$) the corresponding resonance frequencies, and $g_j$ the vacuum optomechanical coupling rates.
The laser driving the cavity is modelled by adding $i\hbar[\mathcal{E}_{\text{drive}}(t)\hat{a}^{\dagger}-\mathcal{E}_{\text{drive}}^*(t)\hat{a}]$ to the Hamiltonian,
where we assume $\mathcal{E}_{\text{drive}}(t)=\mathcal{E}_{0}e^{i\omega_{\text{L}}t}\mathcal{T}(t)$ and $\mathcal{T}(t)$ implements optical modulation.
Appending bath degrees of freedom and tracing them out~\cite{Aspelmeyer2014} yields quantum Langevin equations. These are separable into mean field and fluctuation components  ($\hat{a}(t)e^{i\omega_{\text{L}}t+i\phi_0}=\alpha(t)+\hat{\mathfrak{a}}(t)$ and $\hat{b}_j(t)=\beta_j(t)+\hat{\mathfrak{b}}_j(t)$), with mean fields obeying
\begin{align}
\dot{\alpha}&=\biggl\{-i\biggl[\Delta-\sum\limits_{j=1}^Ng_j\text{R}(\beta_j)\biggr]-\frac{\kappa}{2}\biggr\}\alpha+\mathcal{E}_0\mathcal{T}e^{-i\phi_0}, \nonumber \\
\dot{\beta}_j&=-\bigg(i\Omega_{j}+\frac{\Gamma_j}{2}\bigg)\beta_j+ig_j|\alpha|^2,
\label{eq:MeanFieldEvolution}
\end{align}
and linearized fluctuation dynamics
\begin{align}
\dot{\hat{\mathfrak{a}}}&=-\bigg(i\Delta+\frac{\kappa}{2}\bigg)\hat{\mathfrak{a}}+i\sum\limits_{j=1}^Ng_j[\alpha\mathfrak{R}(\hat{\mathfrak{b}}_j)+\hat{\mathfrak{a}}\text{R}(\beta_j)]+\sqrt{\kappa}\hat{\mathfrak{a}}_{\text{in}}, \nonumber \\
\dot{\hat{\mathfrak{b}}}_j&=-\bigg(i\Omega_{j}+\frac{\Gamma_j}{2}\bigg)\hat{\mathfrak{b}}_j+ig_j(\alpha^*\hat{\mathfrak{a}}+\alpha\hat{\mathfrak{a}}^{\dagger})+\sqrt{\Gamma_j}\hat{\mathfrak{b}}_{j,\text{in}}.
\end{align}
Here, the optical field is considered in a rotating frame with respect to the central laser frequency $\omega_{\text{L}}$ such that $\Delta=\omega_{\text{op}}-\omega_{\text{L}}$ denotes the detuning of the central laser frequency from optical resonance. Additionally, we introduce the optical (mechanical) decay rate $\kappa$ ($\Gamma_j$), input noise operators $\hat{\mathfrak{a}}_{\text{in}}$ ($\hat{\mathfrak{b}}_{j,\text{in}}$), $\mathfrak{R}(\hat{o})=\hat{o}+\hat{o}^{\dagger}$, and $\text{R}(z)=z+z^*$. For a periodic modulation $\mathcal{T}(t)=\sum_k\mathcal{T}_ke^{-ik\Omega_{\text{mod}}t}$ ($k\in\mathds{Z}$), the equation for the mean optical field $\alpha$ inherits the periodicity and admits a Floquet ansatz~\cite{Malz2016,Pietikainen2020}. Hence, we express $\alpha$ as a truncated Fourier series $\alpha(t)=\sum_n\alpha_ne^{-in\Omega_{\text{mod}}t}$ with $n\in\{-D,...,D\}$ and find that \eref{MeanFieldEvolution} reduces to the dynamical system
\begin{align}
\dot{\alpha}_m=&\mathcal{E}_0\mathcal{T}_m-\tilde{\chi}^{-1}_{\text{cav},m}\alpha_{m} +\sum_{(p,q)}\chi^{-1}_{\text{cub},q}\alpha_{p}\alpha_{p-q}^*\alpha_{m-q},
\end{align}
where $p\in\{-D,...,D\}$, $q\in\{-D+p,...,D+p\}$, and the solutions of the mechanical mean fields $\beta_j(t)$ follow from their solutions in Fourier space. Here, we defined $\tilde{\chi}^{-1}_{\text{cav},m}=i(\Delta-m\Omega_{\text{mod}})+\frac{\kappa}{2}$, and $\chi^{-1}_{\text{cub},q}=\sum_{j}\chi^{-1}_{\text{OM},j}(q\Omega_{\text{mod}})$, with $\chi^{-1}_{\text{OM},j}(\omega)/g_j^2=[i(\omega-\Omega_j)-\frac{\Gamma_j}{2}]^{-1}-[i(\omega+\Omega_j)-\frac{\Gamma_j}{2}]^{-1}$. We implement laser intensity modulation by $\mathcal{T}_0=[1-i\mathcal{J}_0(d)]/2$, $\mathcal{T}_{\pm 1}=-\mathcal{J}_1(d)/2$, where $\mathcal{J}_m$ denotes the $m$-th Bessel function of the first kind and $d$ the modulation depth (see Supplemental Material). Employing the steady state $\bar{\alpha}_m$, which has to be found numerically beyond $D=0$ by solving $\dot{\alpha}_m=0$, turns the dynamics of fluctuation components $\hat{\mathfrak{a}}$ and $\hat{\mathfrak{b}}$ into a periodic system that can be treated with Floquet techniques~\cite{Malz2016}:
\begin{align}
\dot{\hat{\mathfrak{a}}}^{(m)}=&-\tilde{\chi}_{\text{cav},m}^{-1}\hat{\mathfrak{a}}^{(m)}-\sum\limits_{(p,q)}\sum\limits_{j=1}^N \chi_{\text{OM},jq}^{-1}\bar{\alpha}_p\bar{\alpha}^{*}_{p-q}\hat{\mathfrak{a}}^{(m-q)} \nonumber \\
&+\sum\limits_{n=-D}^D\sum\limits_{j=1}^N ig_j\bar{\alpha}_{-n}\mathfrak{R}(\hat{\mathfrak{b}}_j^{(m-n)})+\sqrt{\kappa}\hat{\mathfrak{a}}_{\text{in}}^{(m)}, \nonumber \\
\dot{\hat{\mathfrak{b}}}_j^{(m)}=&-\tilde{\chi}_{\text{me},mj}^{-1}\hat{\mathfrak{b}}^{(m)}_j+ig_j\sum_{n=-D}^{D}\bigl[\bar{\alpha}_{-n}^*\hat{\mathfrak{a}}^{(m-n)}+\bar{\alpha}_n\hat{\mathfrak{a}}^{\dagger(m-n)}\bigr]\nonumber \\
&+\sqrt{\Gamma_j}\hat{\mathfrak{b}}_{j,\text{in}}^{(m)}.
\end{align}
 Using the input--output relations for the relevant contributions of the optical field $\hat{\mathfrak{a}}_{\text{out}}(\omega)=\hat{\mathfrak{a}}^{(0)}_{\text{in}}(\omega)-\sqrt{\kappa}\hat{\mathfrak{a}}^{(0)}(\omega)$ with input noise obeying $\langle\hat{\mathfrak{u}}^{(m)}_{\text{in}}(\omega)\hat{\mathfrak{w}}^{\dagger(n)}_{\text{in}}(\omega')\rangle=\delta(\omega-\omega')\delta_{\mathfrak{u}\mathfrak{w}}\delta_{mn}(\mathfrak{n}_{\text{th}}^{\mathfrak{u}}+1)$ yields the stationary power spectral density of the output field consisting of a noise floor $\tilde{S}$ and multiple Lorentzian peaks proportional to $\mathfrak{n}_{\text{th}}^{\mathfrak{b}_j}\equiv \bar{n}_j$
\begin{align}
S(\omega)=\tilde{S}+\sum\limits_{p,j}\frac{\kappa g^2_j|\bar{\alpha}_{p}|^2\Gamma_j\bar{n}_j}{\Big[(\omega-\tilde{\Delta})^2+\frac{\tilde{\kappa}^2}{4}\Big]\Big[(\omega-\Omega_{jp})^2+\frac{\Gamma_j^2}{4}\Big]}
\end{align}
located at $\Omega_{jp}=\Omega_j+p\Omega_{\text{mod}}$ and filtered by the cavity density of states. This is of Lorentzian form, with frequency-independent effective detuning $\bar{\Delta}=\Delta+\sum_{j,p}\frac{2g_j^2|\bar{\alpha}_p|^2}{\Omega_j}$, due to static radiation pressure, modified to $\tilde{\Delta}(\omega)=\bar{\Delta}+\sum_{j,p}|\bar{\alpha}_p|^2\text{Im}(\chi^{-1}_{\text{OM},j}(\omega+p\Omega_{\text{mod}}))$ and effective linewidth $\frac{\tilde{\kappa}}{2}(\omega)=\frac{\kappa}{2}+\sum_{j,p}|\bar{\alpha}_p|^2\text{Re}(\chi^{-1}_{\text{OM},j}(\omega+p\Omega_{\text{mod}}))$.

\emph{Stability analysis.}---To evaluate the mechanical stability, we eliminate the optical field fluctuation operator $\hat{\mathfrak{a}}^{(0)}$ and analyze its effect on the mechanical Floquet modes
\begin{align}
\hat{\mathfrak{b}}&_j^{(m)}[\tilde{\chi}^{-1}_{\text{me},mj}-i\omega]=\sqrt{\Gamma_j}\hat{\mathfrak{b}}_{j,\text{in}}^{(m)}-\sum_{l,p}\sigma^{(m)}_{jlp}(\omega)\mathfrak{R}(\hat{\mathfrak{b}}_l^{(p)}) \nonumber \\
&+i\sqrt{\kappa}g_j\bigg[\frac{\bar{\alpha}_{-m}^{*}\hat{\mathfrak{a}}_{\text{in}}^{(0)}}{i(\bar{\Delta}-\omega)+\frac{\kappa}{2}}+\frac{\bar{\alpha}_{m}\hat{\mathfrak{a}}_{\text{in}}^{(0)\dagger}}{-i(\bar{\Delta}+\omega)+\frac{\kappa}{2}}\bigg],
\end{align}
 where $\tilde{\chi}_{\text{me},mj}^{-1}=i(\Omega_j-m\Omega_{\text{mod}})+\frac{\Gamma_j}{2}$. The Floquet modes are coupled via the contributions
\begin{align}
\sigma_{jlp}^{(m)}(\omega)=\frac{g_jg_l\bar{\alpha}^{*}_{-m}\bar{\alpha}_{p}}{i(\bar{\Delta}-\omega)+\frac{\kappa}{2}}-\frac{g_jg_l\bar{\alpha}_{m}\bar{\alpha}^{*}_{p}}{-i(\bar{\Delta}+\omega)+\frac{\kappa}{2}}.
\end{align}
Without periodic drive ($m\equiv p \equiv 0$), the stationary mechanical spectra are $S_{\hat{\mathfrak{b}_j}}(\omega)=\tilde{S}_{\hat{\mathfrak{b}_j}}+\Gamma_j\bar{n}_j[(\Omega'_j-\omega)^2+\Gamma'^2_j/4]^{-1}$
with modified mechanical frequencies and decay rates~\cite{Genes2009,Karuza2012}
\begin{align}
\Omega'_j(\omega)&=\Omega_j\sqrt{1-\frac{\bar{\Delta} g_j^2|\bar{\alpha}_0|^2[\frac{\kappa^2}{4}-\omega^2+\bar{\Delta}^2]}{\Omega_j[\frac{\kappa^2}{4}+(\omega-\bar{\Delta})^2][\frac{\kappa^2}{4}+(\omega+\bar{\Delta})^2]}}, \nonumber \\
\Gamma'_j(\omega)&=\Gamma_j+\frac{4\kappa g_j^2|\bar{\alpha}_0|^2\bar{\Delta}\omega}{[\frac{\kappa^2}{4}+(\omega-\bar{\Delta})^2][\frac{\kappa^2}{4}+(\omega+\bar{\Delta})^2]}.
\label{eq:selfenergy}
\end{align}
These expressions allow assessing the stability of the mechanical oscillators: The mechanical decay rate $\Gamma'_j$ is composed of decay of phonons into the bath $\Gamma_j$ which can be counteracted by the stimulated emission process for blue-detuned driving ($\bar{\Delta}<0$). In the stimulated process for the respective oscillator $\omega=\Omega_j$, a cavity photon with excess energy and a stimulating phonon are converted into a resonant photon and two coherent phonons of this mode. Its rate overcoming the decay rate $\Gamma_j$ indicates the onset of self-sustained oscillation. 

In the presence of the periodic drive, there are additional contributions that modify the mechanical decay rates
\begin{align}
\Gamma''_j(\Omega_j)=\Gamma'_j-\sum\limits_{l,p}\text{Re}\bigg[\frac{(1-\delta_{jl}\delta_{p0})\sigma_{jlp}^{(0)}\sigma_{jl0}^{(p)}}{i(\Omega_l-\Omega_{jp})+\frac{\Gamma_l}{2}+\sigma_{jlp}^{(p)}}\bigg],
\label{eq:twomodedamping}
\end{align}
which can become prominent if the modulation frequency is tuned to the difference of distinct mechanical frequencies $\Omega_{\text{mod}}=|\Omega_j-\Omega_l|$ with $j \ne l$ for low modulation depths ($|\bar{\alpha}_{\pm1}|^{2}\ll |\bar{\alpha}_{0}|^{2}$). These additional contributions can be interpreted as the stimulated emission of a cavity photon and a phonon creating a coherent phonon in a different mode (see Supplemental Material). This process can act as a seed of phase-locking between the two mechanical modes, and cause a mode that exceeds threshold to stimulate simultaneous lasing of the other mode, circumventing the prohibitive effect of gain saturation. 

To verify the existence of a multimode phonon lasing state, we conduct a numerical simulation of the It\^o stochastic differential equation corresponding to \eref{MeanFieldEvolution} as depicted in \fref{Numerics1}. 
We employ the Euler--Maruyama scheme~\cite{Kloeden1992}, adding Gaussian noise terms $\xi_{r}(t)$ with zero mean $\langle\xi_{s}(t)\rangle=0$ and time correlation $\langle\xi_{r}(t)\xi_{s}(t')\rangle=\delta_{rs}\lambda_s\delta(t-t')$ for all $2(N+1)$ variables $r,s \in \{\text{Re}(\alpha),\text{Im}(\alpha),\text{Re}(\beta_j),\text{Im}(\beta_j)\}$ with the variance of the Gaussian noise $\lambda_s$ (see Supplemental Material). We choose an instructive set of parameters for two mechanical modes ($N=2$, $\Omega_1=5.3$, $\Gamma_1/\Omega_1=0.16$, $g_1=0.80$, $\Omega_2=7.1$, $\Gamma_2/\Omega_2=0.10$, $g_2=1.1$) and the optical cavity ($\Delta=-6.1$, $\kappa=3$). Driving this system with $\mathcal{E}_0=8.9$, leading to $\bar{\Delta}\approx -\Omega_1$, while modulating with depth $d=0.08$ for various modulation frequencies $\Omega_{\text{mod}}$ reveals the effect of intensity modulation. \Fref{Numerics1}(b) shows that for off-resonant intensity modulation ($\Omega_\text{mod}\ll\Omega_2-\Omega_1$) the mechanical modes evolve in a thermal state, as the system is just below instability threshold for the chosen power. For modulation closer to resonance, one of the two modes ($\beta_1$) transitions into self-sustained oscillations, with high amplitude and narrow spectrum as expected for a lasing process (\fref{Numerics1}(c)). This can happen because of the additional negative damping contribution in Eq. (\ref{eq:twomodedamping}). That extra modulation-induced process involves the other mechanical mode ($\beta_2$), as evidenced by the fact that $\beta_2$ is coherently driven at frequencies $\Omega_1$ and $\Omega_1+\Omega_\text{mod}$, while it still does not show significant narrowing (magenta spectrum in \fref{Numerics1}(c)). This changes when the modulation frequency is approximately the difference of the mechanical frequencies (\fref{Numerics1}(d)), and the multimode process gain is maximal: Then, both mechanical modes undergo coherent oscillation at distinct frequencies. We coin this operation multimode oscillation (MMO) since \emph{both} peaks in the mechanical spectra are described by Lorentzians with decreased linewidth and not by the sum of the thermal Lorentzian and additional peaks, as in \fref{Numerics1}(c).

\begin{figure}
\includegraphics[width=0.45\textwidth,angle=0]{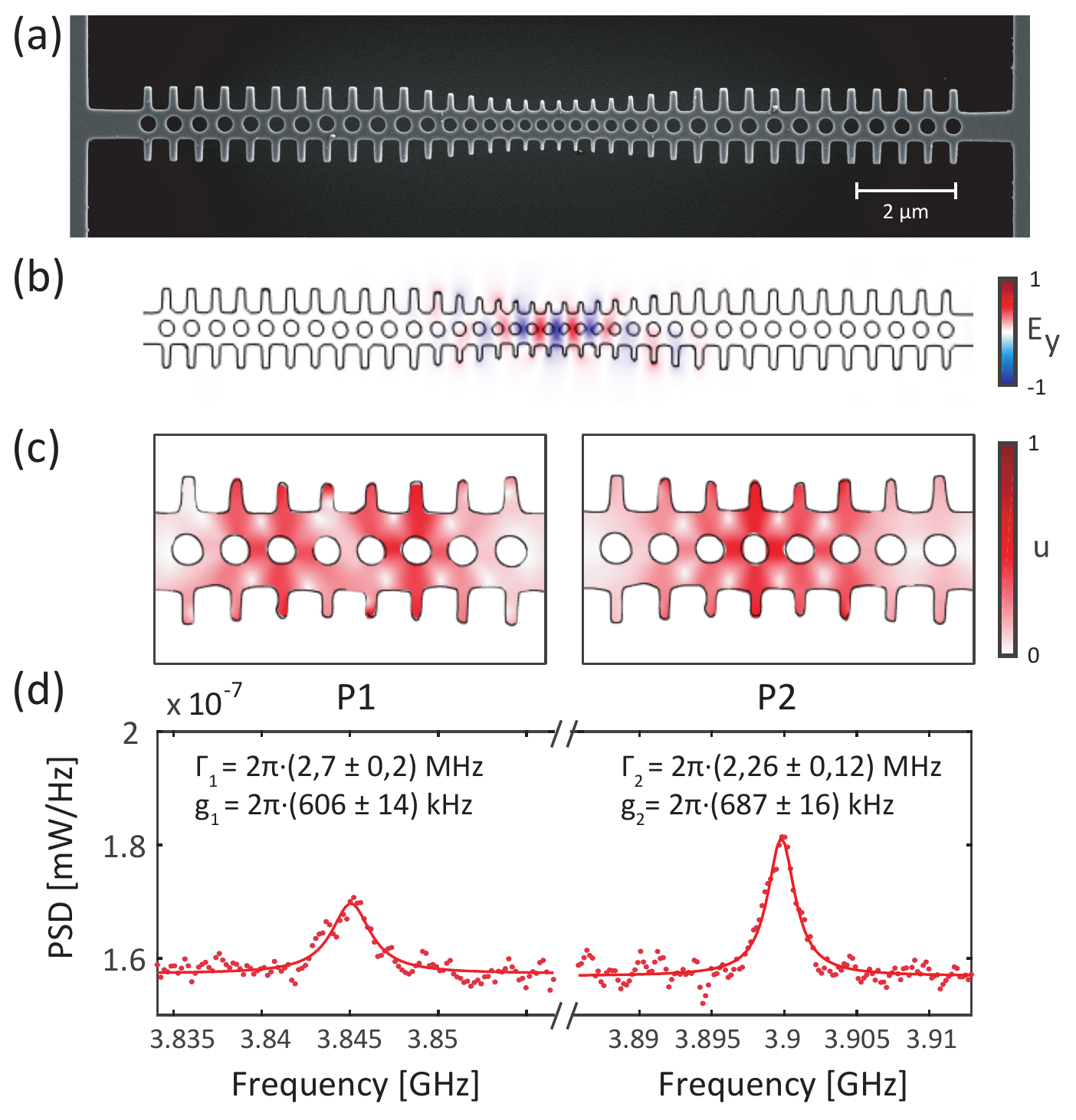}
\caption{(a) Scanning electron microscope image of the fabricated optomechanical cavity. (b) Simulated electric field pattern of the localized optical mode in the fabricated cavity. The mode has a measured resonance wavelength $\lambda_{r}$=(1527.4$\pm$0.2)~nm with a loaded optical quality factor $Q_{o}$=(1.32$\pm$0.1)$\times$10$^{4}$ and an overall decay rate $\kappa/2\pi$=(14.88$\pm$0.08) GHz. (c) Calculated mechanical displacement profiles of the two mechanical modes P1 and P2. (d) Measured power spectral density of the thermally-transduced mechanical modes P1 and P2. These allow determination of resonance frequencies $\Omega_{1}/2\pi$=3.845 GHz and $\Omega_{2}/2\pi$=3.899 GHz, linewidths $\Gamma_{1}/2\pi$=(2.7$\pm$0.2) MHz and $\Gamma_{2}/2\pi$=(2.26$\pm$0.12) MHz, and optomechanical coupling rates  $g_{1}/2\pi$=(606$\pm$14) kHz and $g_{2}/2\pi$=(687$\pm$16) kHz.}
\label{fig:F1}
\end{figure}

\emph{Experimental multimode phonon lasing.}---In our experiments to confirm the predicted MMO state, we use the 1D silicon optomechanical crystal cavity depicted in \fref{F1}(a). It supports a high-quality optical mode (\fref{F1}(b)) that allows phonon lasing at frequencies around 4 GHz under blue-detuned laser driving~\cite{Mercade2020}. Interestingly, this system hosts two mechanical modes within the beam's phononic bandgap, which correspond to oscillations of the lateral corrugations~\cite{Oudich2014}. The mechanical displacement of these modes, labeled `P1' and `P2', is depicted in \fref{F1}(c), simulated using the shape retrieved from electron microscopy.
Figure~\ref{fig:F1}(d) shows both modes' thermomechanical spectra transduced at low power. All measurements were performed at room temperature and atmospheric pressure, coupling light to and from the cavity with a dimpled fiber taper (see more experimental details in the Supplemental Material). Notably, the coupling rate and damping of both mechanical modes are quite similar, leading to similar cooperativity $C_j=4g_j^{2}\bar{n}_\text{c}/(\Gamma_j\kappa)$, with $\bar{n}_\text{c}$ the intracavity photon number: $C_1/\bar{n}_\text{c}=(3.7\pm0.3)\times10^{-5}$ for mode P1 and  $C_2/\bar{n}_\text{c}=(5.6\pm0.4)\times10^{-5}$ for mode P2. 

\begin{figure}
\includegraphics[width=0.45\textwidth,angle=0]{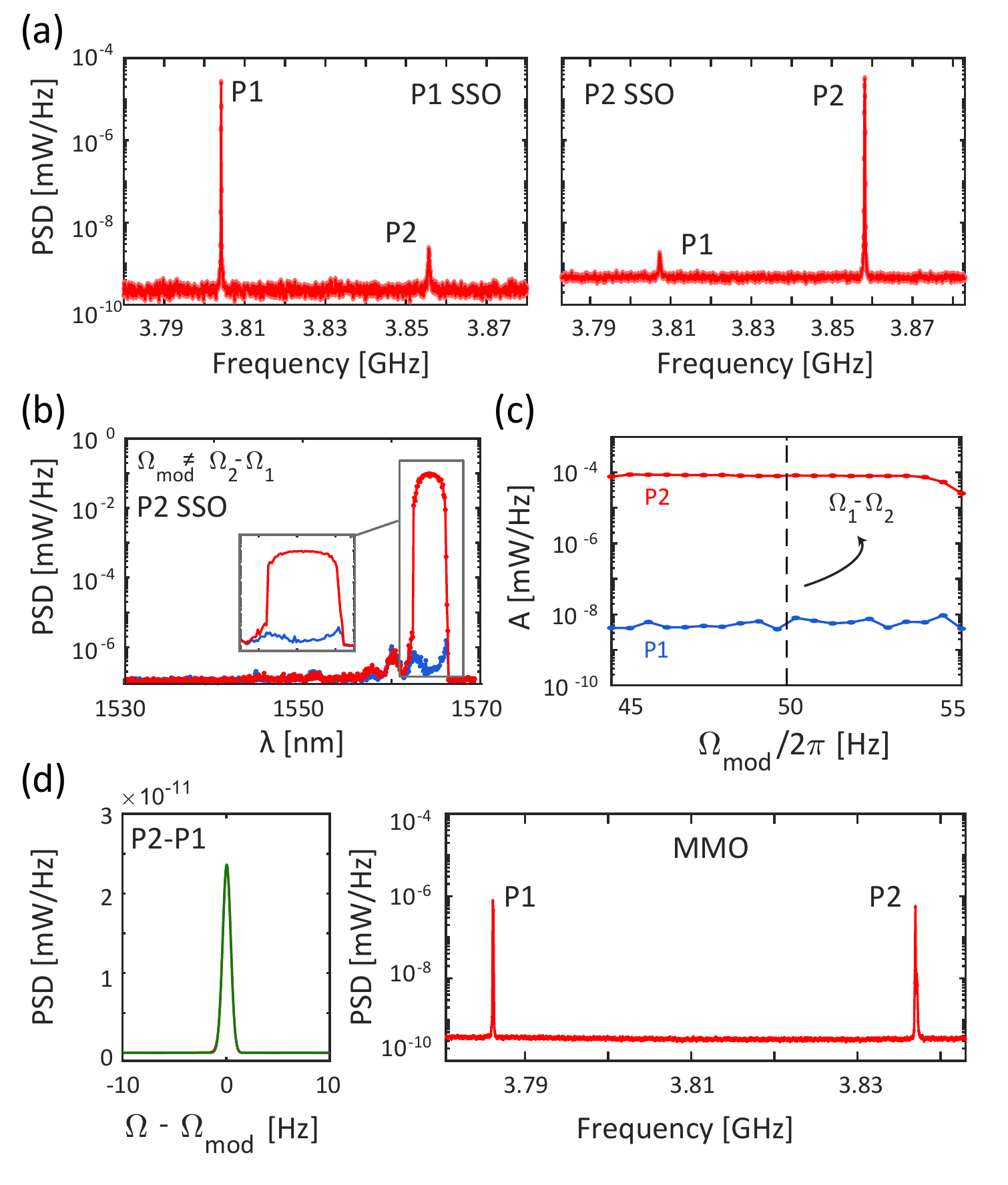}
\caption{From single mode to multimode phonon lasing. (a) Self-oscillation spectra of P1 (left) and P2 (right), without external modulation. (b) P2 SSO excitation through a wavelength scan with a blue-detuned laser $\Omega_{\text{mod}}$ $\neq$ $\Omega_{2}-\Omega_{1}$. (c) Mechanical mode amplitude evolution in a modulation frequency scan around the difference frequency when P2 is self-oscillating. (d) Multimode lasing under an input modulation $\Omega_{\text{mod}}$ = $\Omega_{2}-\Omega_{1}$. (Left) Difference tone obtained by mixing the lasing mechanical modes in the microwave domain. (Right) Spectrum of the two mechanical modes when lasing simultaneously (MMO state).}
\label{fig:F2}
\end{figure}

Because of this similarity, we observe that either mode can be individually driven to a single-mode self-sustained oscillation (SSO) state under blue-detuned driving, as shown in \fref{F2}(a). 
The choice of the lasing state depends on fine experimental conditions, including the initial (random) thermal state of each mode. 
Figure~\ref{fig:F2}(b) depicts the intensity of both modes while the wavelength of the laser (input power $P_{in}=3.16$~mW) is continuously increased towards the SSO regime. 
We see that once one mechanical mode (P2) starts lasing, the other mechanical mode (P1) is damped, evidencing gain suppression \cite{Kemiktarak2014}. Following the theoretical predictions, we then modulate the laser intensity at frequency $\Omega_{\text{mod}}$ (with $d=0.18$). 
However, starting from P2 in an SSO state and stepping the modulation frequency across resonance $\Omega_{\text{mod}}=\Omega_2-\Omega_1$ does not activate the MMO state, as the thermomechanical Lorentzian of P1 remains unaffected (see \fref{F2}(c) and Supplemental Material). 
Instead, the MMO regime is reached when the intermodal coupling is established already before threshold is reached, allowing the Floquet modes to cross threshold simultaneously. We experimentally achieve this by implementing the modulation fixed to the measured difference frequency, and then tuning a far-blue-detuned laser towards cavity resonance. 
The resulting MMO state is shown \fref{F2}(d) (right): both modes are simultaneously in the self-sustained regime. The left panel shows the beatnote of the two lasing tones, measured after mixing them and filtering out the modulation at $\Omega_\text{mod}$ (see Supplemental Material). 
Its extremely narrow linewidth, much smaller than the linewidths of the individual tones and only limited by the 1 Hz resolution bandwidth of the spectrum analyzer, confirms the phase locking of the two lasing phonon modes, whose phase fluctuations are identical.

\begin{figure}
\includegraphics[width=0.45\textwidth,angle=0]{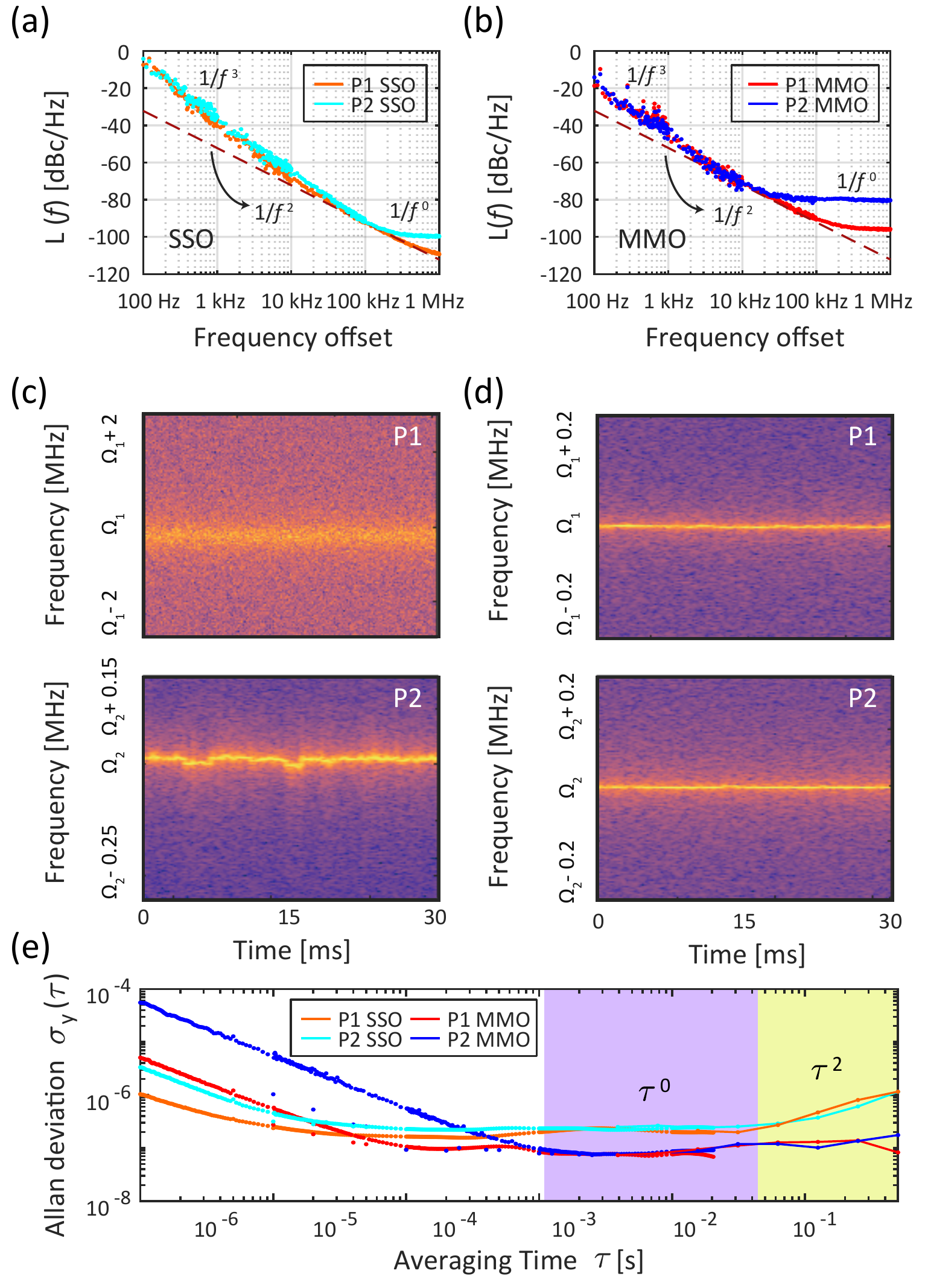}
\caption{Phase noise and stability behaviour. Phase noise of P1 (blue) and P2 (red) in the SSO (a) and MMO regimes (b). Time evolution of the recorded spectra for the P1 (top) and P2 (bottom) modes for SSO lasing in mode P2 (c), and for MMO lasing (d). (e) Allan deviation for P1 and P2 in the SSO and MMO states.}
\label{fig:F4}
\end{figure}

\emph{Phase noise and stability analysis.}--- To characterize the linewidth and stability of the oscillators, we analyze their phase noise $L(f)$ in the SSO and MMO states in Figs.~\ref{fig:F4}(a,b). 
In both cases we observe different noise contributions that can be expected in optomechanical oscillators: white phase ($1/f^{0}$), white frequency (random phase walk, $1/f^{2}$), and flicker frequency ($1/f^{3}$) noise~\cite{Mercade2020}, in good agreement with Leeson's model~\cite{Rubiola}.
A phase noise of ($-65\pm3$) dBc/Hz at 10 kHz is measured for the SSO state, on par with other optomechanical microwave oscillators~\cite{TAL11-OE,BHA12,Luan2014,GHOR19-APLPhot,Mercade2020}. Although white phase and flicker frequency noise are dominating, we estimate a Lorentzian linewidth of maximally $\sim40$~Hz from the fitted contribution of the random phase walk ($1/f^2$).

Interestingly, the phase noise of the MMO state is ($-70\pm4$) dBc/Hz at 10 kHz; thus smaller than that in the SSO regime for this and all smaller offset frequencies. This reduction of fluctuations is also very clear from the time evolution of the spectra in the SSO (P2) and MMO states, presented in Fig. \ref{fig:F4}(c,d). The observed jitter of the lasing frequency of P2 in panel (c) could stem from thermal, optical power, and fiber taper fluctuations influencing the optical spring effect, as \eref{selfenergy} implies.
In the MMO lasing state (panel (d)), we observe a significantly enhanced stability of both lasing modes.  

Figure~\ref{fig:F4}(e) shows the Allan deviation $\sigma(\tau)$~\cite{Allan1997}, obtained from combining the phase noise measurements in \fref{F4}(a,b) and the spectral time traces for longer acquisition times (up to 1 s).  
For low averaging times $\tau$, the main differences between the Allan deviation stem from the white phase noise background, which is related to measurement sensitivity and signal amplitude. However, for longer averaging times (purple shaded area), where the dominant fluctuations are flicker frequency noise ($\sigma(\tau)\propto\tau^{0}$), the MMO is significantly more stable than the SSO state. This is in agreement with the obtained root-mean-square jitter, which reduces from $220\pm5$~ps in both SSO states to $76\pm5$~ps in the MMO state (see Supplemental Material). Also for high averaging times with a frequency drift contribution ($\sigma\propto\tau^{2}$, yellow shaded area), the multimode stability is superior. It should be noted that this improvement cannot be ascribed to locking to the external modulation, as that is far off-resonant. Instead, it could be related to the MMO regime being sustained at significantly reduced intracavity power, as suggested by the relative heights in \fref{F2}, thus reducing the influence of optical fluctuations on the optomechanical spring effect~\cite{ANT12-NCOMMS}. Moreover, it is known that synchronized oscillators become less prone to fluctuations as their effective mass is increased \cite{Lipson2015, CROS12-PRE}. This effect could very well be in play here, with an associated 3 dB reduction of phase noise expected when the two oscillators are coupled. Further studies should be undertaken to reveal the various contributions and the application potential of the increased stability in the Floquet lasing regime.

\emph{Conclusion.}---Our investigation shows that nonlinear dynamics of Floquet modes enable multimode phonon lasing in optomechanical systems. The analytical argument based on higher-order cross-mode corrections of the self-energy for intensity modulation at the difference of mode frequencies is underpinned by numerical simulations as well as experimental demonstration. We show how the multimode lasing state can be reached experimentally, and reveal that its long-term stability is improved. The multimode cavity used in our experiments paves the way towards to integration of multiple coexisting GHz mechanical modes. 
The concepts we establish can be used to study the emergent physics of many-mode self-oscillating optomechanical systems, in a fully controllable fashion.


\begin{acknowledgments}
The authors thank Javier del Pino for useful discussions. This work is supported by the European Union's Horizon 2020 research and innovation programme under Grant Agreements Nos.\ 732894 (FET Proactive HOT), 713450 (FET-Open PHENOMEN), and 945915 (SIOMO). It is part of the research programme of the Netherlands Organisation for Scientific Research (NWO). A.M.\ acknowledges funding from Generalitat Valenciana under grants PROMETEO/2019/123, BEST/2020/178, and IDIFEDER/2018/033. E.V. acknowledges support from the European Research Council (ERC Starting Grant No. 759644-TOPP). L.M. and A.M. thank AMOLF for the hospitality during their research visits.
\end{acknowledgments}

\appendix

%

\clearpage
\pagebreak
\renewcommand{\theequation}{S\arabic{equation}}
\renewcommand{\thefigure}{S\arabic{figure}}
\onecolumngrid
\begin{center}
    \textbf{\large Supplemental Material}\\
    Floquet phonon lasing in multimode optomechanical systems
\end{center}

\setcounter{equation}{0}
\setcounter{figure}{0}
\setcounter{table}{0}
\setcounter{page}{1}
\makeatletter
\vspace{0.9cm}
\twocolumngrid

\renewcommand{\thesection}{\arabic{section}}

\section*{I. Transfer characteristic of intensity modulation}
\begin{figure}
\begin{center}
\includegraphics[width=0.5\textwidth]{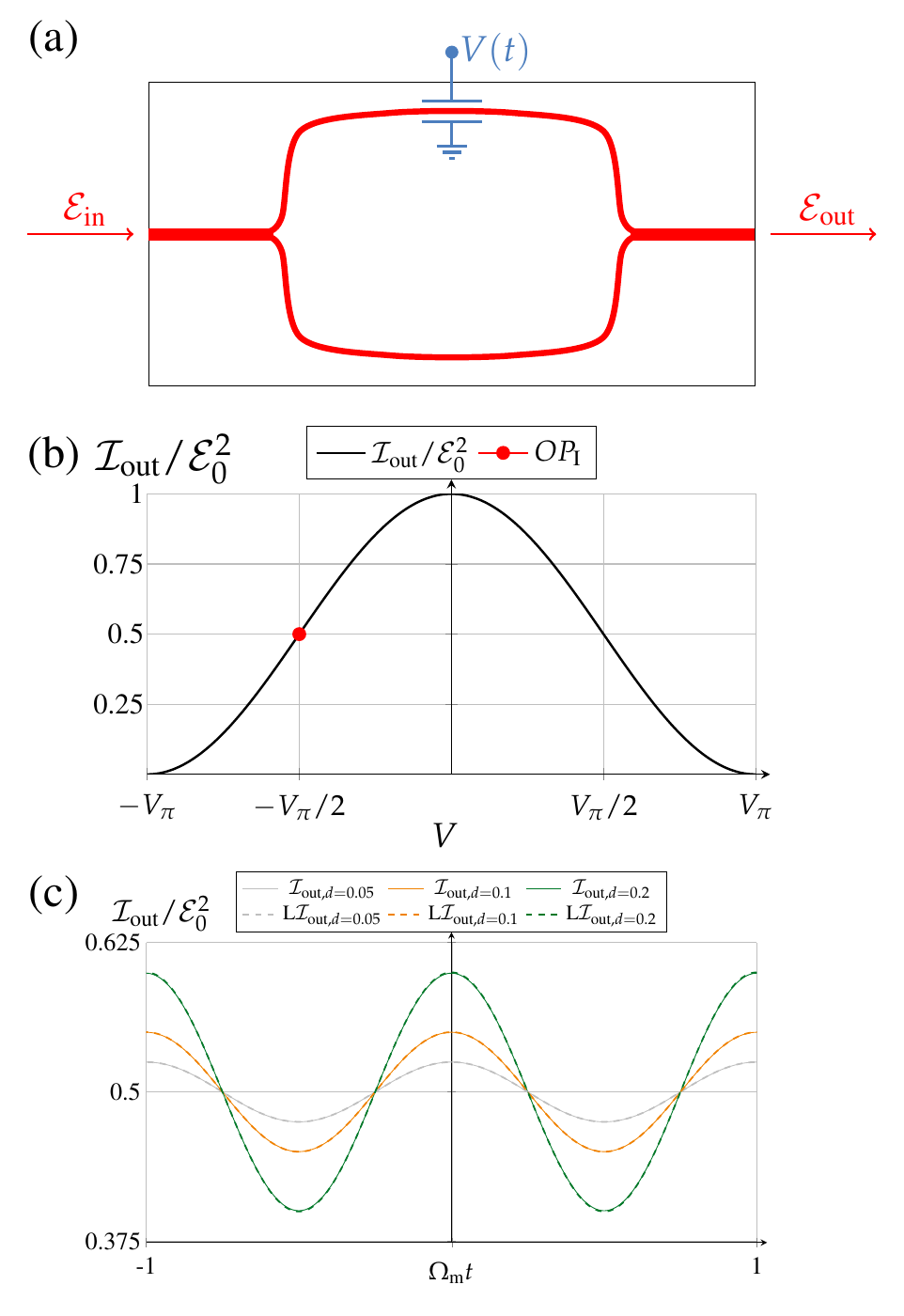}%
\end{center}
\caption{Schematic description of a Mach--Zehnder intensity modulator. (a) Schematic diagram of a Mach--Zehnder interferometer and an electro-optic phase modulator implemented in one optical waveguides. (b) Resulting intensity transfer characteristic as a function of the applied voltage $V$ on the electro-optic phase modulator. (c) Comparison of the intensity transfer function (solid) with the lowest order approximation (dashed) for periodic modulation with low modulation depth $d$.}
\label{fig:mod}
\end{figure}

The optical intensity modulation underlying the theoretical description is based on the imbalanced single-drive Mach--Zehnder modulator depicted in Fig.~\ref{fig:mod}(a). Our discussion follows the descriptions of~\cite{Saleh1991,Seimetz2009}. It consists of a Mach--Zehnder interferometer realized with two waveguides and an electro-optic phase modulator implemented in one waveguide. The output field of such a Mach--Zehnder modulator is given by
\begin{align}
\mathcal{E}_{\text{out}}(t)=\frac{\mathcal{E}_{\text{in}}(t)e^{i\phi_0}}{2}\big(1+e^{i\phi_{\text{mod}}(t)}\big)
\end{align}
where $\mathcal{E}_{\text{in}}$ is the input field and $\phi_0$ an input phase offset. The electro-optic phase modulator causes the phase-shift $\phi_{\text{mod}}(t)=\pi V(t)/V_{\pi}$ when a voltage $V(t)$ is applied, where $V_\pi$ is a characteristic of the modulator. The intensity transfer characteristic for $\mathcal{E}_{\text{in}}(t)=\mathcal{E}_0e^{i\omega_Lt}$ is
\begin{align}
\frac{\mathcal{I}_{\text{out}}}{\mathcal{E}_{0}^2}=\frac{2+e^{i\phi_{\text{mod}}(t)}+e^{-i\phi_{\text{mod}}(t)}}{2}=\cos^2\bigg(\frac{\phi_{\text{mod}}(t)}{2}\bigg)
\label{eq:ITan}
\end{align}
and enables its use for intensity modulation at the operating point $OP_{\text{I}}$ for $V(t) \approx -\frac{V_{\pi}}{2}$ as suggested by Fig.~\ref{fig:mod}(b). Assuming sinusoidal modulation $V(t)=-\frac{V_{\pi}}{2}+\frac{d V_{\pi}}{\pi}\cos(\Omega_{\text{mod}}t)$ around voltage $-V_\pi/2$ with modulation depth $d$ we write the transfer function 
\begin{align}
\mathcal{T}(t)=\frac{\mathcal{E}_{\text{out}}(t)}{\mathcal{E}_{\text{in}}(t)}&=\frac{e^{i\phi_0}}{2}\big(1+e^{-\frac{i\pi}{2}+id\cos(\Omega_{\text{mod}}t)}\big).
\end{align}
The field transfer characteristic can be described in terms of the Bessel functions of the first kind $\mathcal{J}_n$ using the Jacobi-Anger expansion
\begin{align}
\frac{\mathcal{T}(t)}{e^{i\phi_0}}=\frac{1-i\mathcal{J}_0(d)}{2}+\sum\limits_{n=1}^{\infty}i^{n+1} \mathcal{J}_n(d)\cos(n\Omega_{\text{mod}}t),
\label{eq:ITJA}
\end{align}
which signifies that applying a low modulation depth $d$ allows to cut off after the first time-dependent contribution. The Fourier components of Eq. (\ref{eq:ITJA}) act as the driving terms $\mathcal{T}_m$ in Eq. (4) of the main text. To ensure the validity of this cut-off, we compare the analytical intensity transfer function in \eref{ITan} with the approximation resulting from truncating Eq. (\ref{eq:ITJA}) after $n=1$ in Fig.~\ref{fig:mod}(c) in the parameter regime used in theory as well as the experiment.

\section*{II. Raman picture description}
\begin{figure}
\begin{center}
\includegraphics[width=0.5\textwidth]{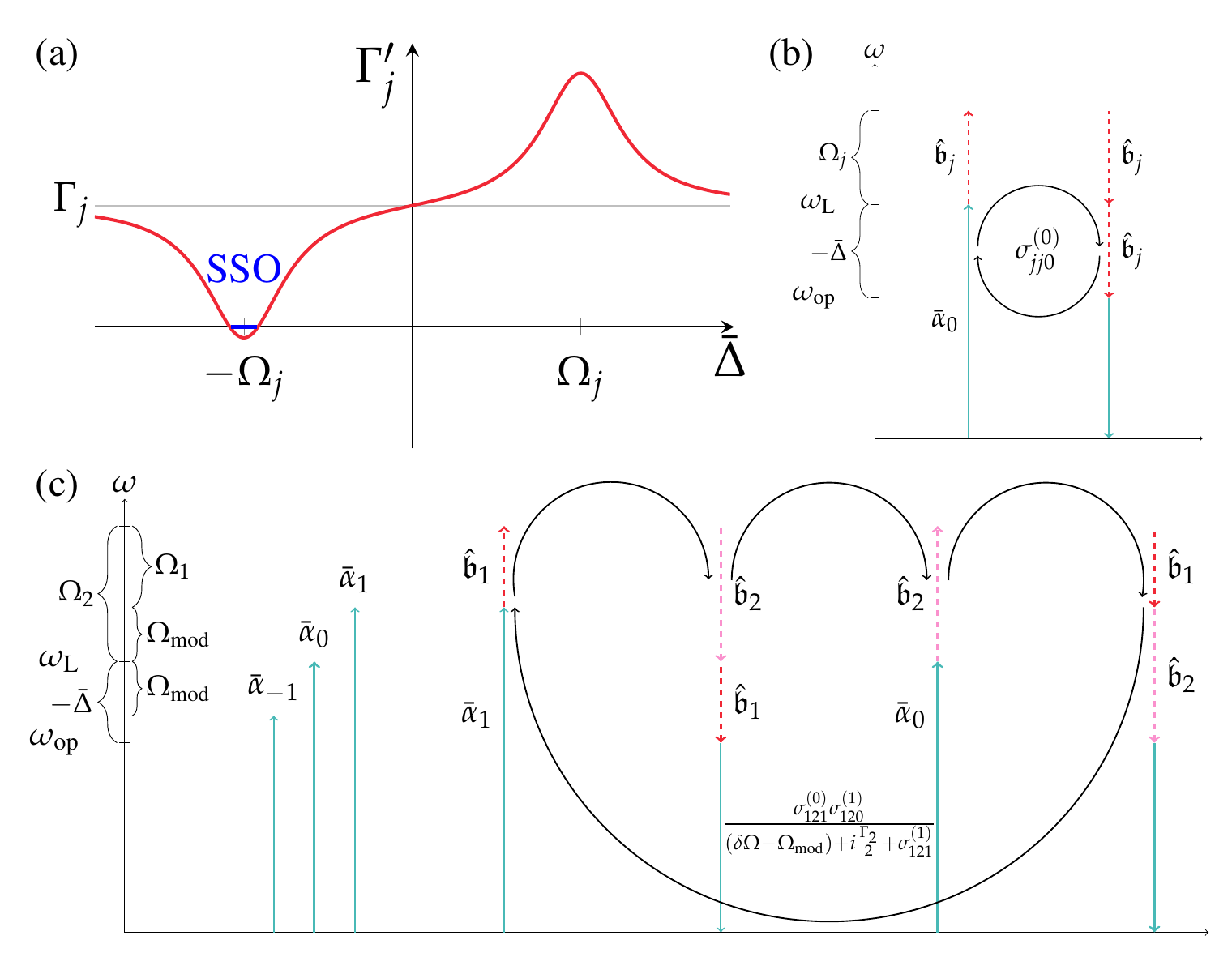}%
\end{center}
\caption{Schematic diagrams of self-sustained oscillation (SSO) and multi-mode oscillation (MMO). (a) Effective phonon decay rate $\Gamma'_j$ of a sideband resolved optomechanical system as a function of the effective detuning $\bar{\Delta}$. The initial decay rate $\Gamma_j$ is corrected by another process which can enhance or decrease the decay. If the drive photons' frequency $\omega_{\text{L}}$ exceed the optical cavity frequency $\omega_{\text{op}}$ by the frequency of a mechanical mode $\Omega_{j}$ the decay rate is minimal, indicating energy conservation as a requirement for stimulated emission (b) Schematic diagram based on energy conservation for self-sustained oscillation. In presence of a noise phonon of mode $\hat{\mathfrak{b}}_j$, a photon with the correct excess energy $\hbar\Omega_j$ can be broken down into a resonant photon and an additional phonon in the mode $\hat{\mathfrak{b}}_j$ sharing the phase of the original phonon. (c) Schematic diagram based on energy conservation for multi-mode oscillation. The intensity modulation generates photons with frequencies diminished ($\bar{\alpha}_{-1}$) or augmented ($\bar{\alpha}_{1}$) by the modulation frequency $\omega_{\text{mod}}$ from the central laser frequency $\omega_{\text{L}}$ taken to be $\omega_{\text{op}}+\Omega_1$. The initial step of the multi-mode lasing process breaks an augmented frequency photon ($\bar{\alpha}_{1}$) in presence of a noise phonon $\hat{\mathfrak{b}}_1$ down into a resonant photon and a phase locked phonon $\hat{\mathfrak{b}}_2$ in mode 2. This step is energy conserved if the modulation frequency is the difference of the mechanical frequencies $\delta\Omega=\Omega_2-\Omega_1$. In the second step, this phase-locked phonon breaks up a photon ($\bar{\alpha}_0$) with the central frequency $\omega_{\text{L}}$ into a resonant photon and a phase locked phonon $\hat{\mathfrak{b}}_1$ in mode 1.}
\label{fig:raman}
\end{figure}

With the aim to provide further insight in the working principle of the stimulated emission processes, we first inspect the effective decay rate without periodic drive $\Gamma'_j(\Omega_j)$ taken from Eq.~(9) of the main text. An example in the resolved sideband regime ($\kappa<\Omega_j$) is shown in Fig.~\ref{fig:raman}(a) which illustrates that the initial phonon decay rate $\Gamma_j$ can be counteracted by a phonon production process that is maximal for an effective detuning $\bar{\Delta}=\omega_{\text{op}}-\omega_{\text{L}}=-\Omega_j$. This means that the phonon production is most effective if drive photons entering the cavity carry the exact amount of excess energy $E=\hbar(\omega_{\text{L}}-\omega_{\text{op}})$ that is associated with one phonon $E_{j}=\hbar\Omega_j$, indicating that energy conservation of incoming and outgoing particles is the requirement for the process. Therefore, we can visualize stimulated emission processes in an energy diagram with arrows pointing upwards as incoming particles and arrows pointing downwards as outgoing particles, as depicted in Fig.~\ref{fig:raman}(b) and (c). The standard stimulated emission process displayed in Fig.~\ref{fig:raman}(b) can be visualized by an incoming photon ($\bar{\alpha}_0$) and a (stimulating) phonon $\hat{\mathfrak{b}}_j$ which can be used to break down the photon into a resonant photon at the optical frequency and two phonons $\hat{\mathfrak{b}}_j$ that inherit the phase from the stimulating phonon. Energy conservation of this process requires that the photon carries the excess energy of one phonon $\bar{\Delta}=-\Omega_j$. 

The collaborative stimulated emission process that is relevant for phase locked multi-mode oscillation in the numerical example is illustrated in Fig.~\ref{fig:raman}(c). This process is associated with the decay rate alteration Eq.~(10) in the main text. In the presence of modulation, the cavity is driven with carrier photons ($\bar{\alpha}_0$) at frequency $\omega_\text{L}$ that carrying excess energy $\bar{\Delta}=-\Omega_1$ as well as by photons in the modulation sidebands whose frequencies $\omega_{\text{L}}\pm\Omega_{\text{mod}}$ are either diminished ($\bar{\alpha}_{-1}$) or augmented ($\bar{\alpha}_1$) from the carrier frequency by the modulation frequency $\Omega_{\text{mod}}$. A new process can be recognized that can be broken down into two subcycles: 
\begin{enumerate}
\item{The sideband photons ($\bar{\alpha}_1$) can be broken down by a stimulating phonon of mode 1 ($\hat{\mathfrak{b}}_1$) into a resonant photon and a new phase-locked phonon of mode 2 ($\hat{\mathfrak{b}}_2$). This process conserves energy if the photon's excess energy $-\hbar(\bar{\Delta}-\Omega_{\text{mod}})$ equals the energy $\hbar\Omega_{2}$ of a phonon in mode 2 which yields a condition for the modulation frequency $\Omega_{\text{mod}}=\Omega_{2}+\bar{\Delta}$. Considering a central detuning $\bar{\Delta}=\omega_\text{op}-\omega_\text{L}=-\Omega_1$ this results in modulating at the difference of the mechanical frequencies $\Omega_{\text{mod}}=\Omega_{2}-\Omega_{1}\equiv \delta\Omega$.} 
\item{This phase-locked phonon of mode 2 can in turn be used in a subsequent step to generate a new phonon of mode 1. This step requires a drive photon ($\bar{\alpha}_0$) with excess energy $-\hbar\bar{\Delta}=\hbar\Omega_1$ and the phase locked phonon in mode 2 ($\hat{\mathfrak{b}}_2$) to break down into a resonant photon and a phase locked phonon in mode 1 ($\hat{\mathfrak{b}}_1$) which can act as a stimulating phonon in the next cycle.} 
\end{enumerate}
In total, we require one drive photon in the optical mode $\bar{\alpha}_0$, one drive photon in the optical mode $\bar{\alpha}_1$, and one stimulating phonon in $\hat{\mathfrak{b}}_1$ (e.g. thermal noise) to generate a new pair of phase locked phonons in the mechanical modes 1 and 2. This process can be attributed to the additional term $\sigma_{121}^{(0)}\sigma_{120}^{(1)}/[(\delta\Omega-\Omega_{\text{mod}})+i(\Gamma_2/2)+\sigma_{121}^{(1)}]$ which appears in Eq. (10) of the main text. The collaborative stimulated emission requires that the driving occurs at the difference frequency and both stimulated emission processes are occuring with maximal rates for the effective detuning $\bar{\Delta}=-\Omega_1$ in this example. Therefore, adjusting the modulation frequency with optimal detuning only enables the collaborative stimulated emission process while the standard stimulated emission is already active. However, adjusting the detuning with an optimal modulation enables both stimulated emission processes simultaneously. We note that similar processes occur starting from a noise phonon in mode 2 at detuning $\bar{\Delta}=-\Omega_2$ involving the other modulation sideband, all resulting in phase-locked generation of phonon pairs in the two modes.

\section*{III. Musical description and aural evidence of multimode lasing}
\begin{figure}
\begin{center}
\includegraphics[width=0.5\textwidth]{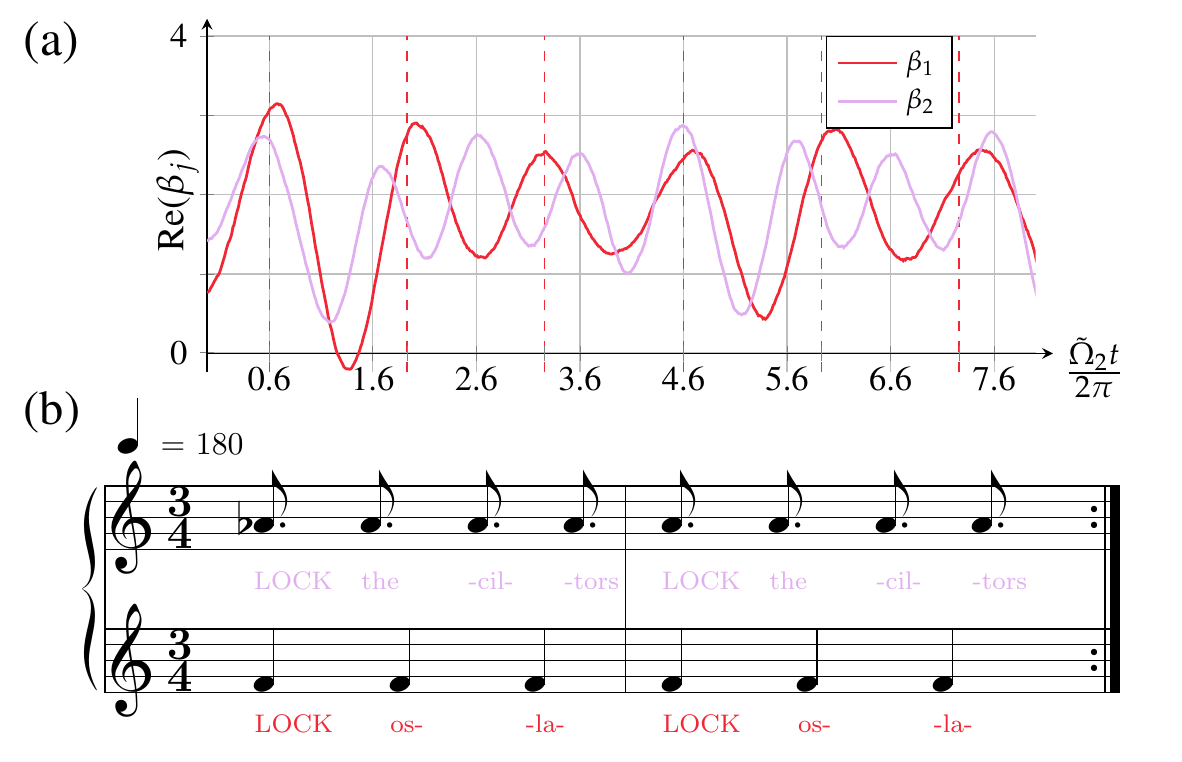}%
\end{center}
\caption{A 4/3 polyrhythm as the musical analogon of multi mode lasing. (a) We show the mechanical traces in the MMO state from Fig.~1(d) of the main text with the main vertical grid lines showing steps of $\tilde{\Omega}_2$ and therefore following the maxima of $\beta_2$ and the red dashed lines showing steps of $\tilde{\Omega}_1$ following the maxima of $\beta_1$. Since the ratio of the frequencies is roughly 4/3 we expect that $\beta_1$ roughly completes 3 oscillations in the time that $\beta_2$ needs for 4 oscillations. (b) Musical notation of the 4/3 polyrhythm in which the lower voice plays three quarters in a three quarter measure such that the notes occur with a frequency of 3 Hz if there are 180 beats per minute. The higher voice plays four dotted eighth notes occuring with a frequency of 4 Hz which fit in the same measure. Additionally the lyrics indicate a mnemonic whose natural rhythm follows the rhythmical pattern. A potential conductor would indicate the beginning of each measure, coinciding with the syllable LOCK in both voices. This means that the conductor gives a sign to the two voices indicating the beginning of the measure occuring with a frequency 1 Hz which is the difference of 4 Hz and 3 Hz. }
\label{fig:Music}
\end{figure}

Another way of appreciating that a modulated drive can lead to phase locked oscillations at distinct frequencies can be drawn from music and more concretely from considering a conductor in charge of leading two groups. The natural frequencies of the mechanical modes in the numerical demonstration were chosen at a ratio $\Omega_2/\Omega_1=1.339\approx 4/3$ such that coherent phase-locked oscillations can be identified with a so-called 4/3 polyrhythm as visualized in Fig.~\ref{fig:Music}. A conductor usually indicates the beginning of a measure which is the smallest cyclic pattern of a musical piece. A measure is subdivided into pulses and here we assume a three quarter measure, i.e. a subdivision into three equal pulses. If a measure takes up one second, then each pulse takes up a third of a second. However, it is possible for a second voice to play four equidistant notes in the same time which are so-called dotted eighths and take up a quarter second each. This fits completely with our example since the two voices repeating notes at a frequency of 3 and 4 Hz are conducted by signs indicating a phase to lock to at the difference of these frequencies, namely at 1 Hz, like the conductor indicating the start of each measure. To validate the existence of multiple distinct frequencies in the oscillations of the mechanical modes, we use the displacement of the numerical simulations as seen in Fig.~1 of the main text to generate sound files. These result in a noisy tone for the thermal state, a single clear tone in the SSO state and in a two-tone chord in the MMO state. Since the natural frequencies are chosen approximately in a 4/3 ratio we expect to hear a so-called suspended fourth chord in the MMO state. We generated the sound files available as part of the supplemental material to certify oneself.

\section*{IV. Numerical analysis procedure and phase diagram}
\begin{figure}
\begin{center}
\includegraphics[width=0.5\textwidth]{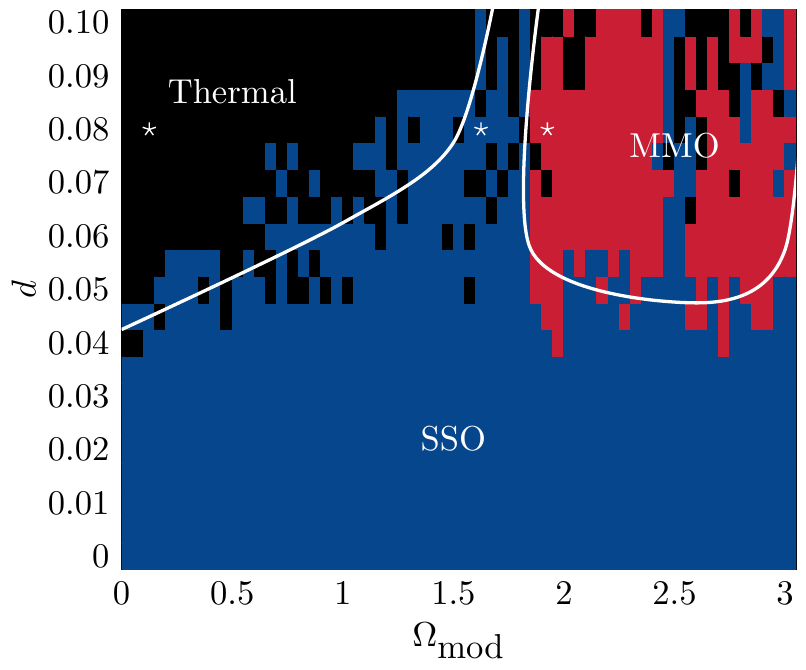}%
\end{center}
\caption{Two-dimensional parameter scan of the modulation depth $d$ and modulation frequency $\Omega_{\text{mod}}$. For low modulation depths we find that the system goes into single mode self-sustained oscillation. Increasing the modulation depth for off-resonant modulation frequencies $\Omega_{\text{mod}}\ll \Omega_2-\Omega_1$, the system is taken below the lasing threshold. For resonant modulation frequencies $\Omega_{\text{mod}}\approx \Omega_2-\Omega_1$ we find multimode self-sustained oscillations. The white stars denote the driving conditions taken for the simulations in Fig.~1 of the main text.}
\label{fig:PhaseD}
\end{figure}

We generate sample trajectories of our model such as those depicted in Fig. 1 of the main text by employing the Euler--Maruyama scheme~\cite{Kloeden1992} for the dynamics of the mean fields
\begin{align}
\dot{\alpha}&=\biggl\{-i\biggl[\Delta-\sum\limits_{j=1}^Ng_j\text{R}(\beta_j)\biggr]-\frac{\kappa}{2}\biggr\}\alpha+\mathcal{E}_0\mathcal{T}e^{-i\phi_0}+\xi_{\alpha}(t), \nonumber \\
\dot{\beta}_j&=-\bigg(i\Omega_{j}+\frac{\Gamma_j}{2}\bigg)\beta_j+ig_j|\alpha|^2+\xi_{\beta_j}(t),
\label{eq:MeanFieldEvolutionstoch}
\end{align}
where we choose the parameters of the two mechanical modes ($N=2$) as described in the main text, namely $\Omega_1=5.3$, $\Gamma_1/\Omega_1=0.16$, $g_1=0.80$, $\Omega_2=7.1$, $\Gamma_2/\Omega_2=0.10$, $g_2=1.1$ as well as the optical cavity $\Delta=-6.1$, and $\kappa=3$. This places the numerical example in the resolved sideband region such that only the collaborative stimulated emission process in Fig. \ref{fig:raman}(c) contributes and the similar process starting from a noise phonon in mode 2 mentioned above is suppressed. We keep all parameters within two orders of magnitude because stiff stochastic differential equations, having parameters varying over several orders of magnitude, cannot easily be simulated numerically. The Gaussian noise terms we employ are described by their statistical momenta, i.e. their mean $\langle\xi_{s}(t)\rangle=0$ taken to be zero thoughout the analysis and time correlation $\langle\xi_{r}(t)\xi_{s}(t')\rangle=\delta_{rs}\lambda_s\delta(t-t')$ for all $2(N+1)$ variables $r,s \in \{\text{Re}(\alpha),\text{Im}(\alpha),\text{Re}(\beta_j),\text{Im}(\beta_j)\}$ with the variance of the Gaussian noise $\lambda_s$ gauging the strength of the random forces. In order to generate realistic initial conditions of the system, we evolve the system starting from rest $\alpha(t=-t_0)=\beta_j(t=-t_0)=0$ and without drive ($\mathcal{E}_0=0$) for an initial period of $t_0=100\cdot \Omega_2/(2\pi) \approx 115$ oscillation periods of $\Omega_2$ emulating cavity shot noise, i.e. $\langle\xi_{r}(t)\xi_{s}(t')\rangle=(\delta_{\text{Re}(\alpha) r}+\delta_{\text{Im}(\alpha) r})\delta_{rs}\delta(t-t')$. Note, that we neglect the equivalent quantum noise of the mechanical oscillator in the ground state because we aim to explore self-oscillating attractors of the nonlinear dynamics with the least energy possible in the system at initial time $t=0$. After the inital procedure to thermalize the system, we then drive with $\mathcal{E}_0=8.9$, leading to $\bar{\Delta}\approx -\Omega_1$, and phonon noise $\langle\xi_{r}(t)\xi_{s}(t')\rangle=0.01\sum_j(\delta_{\text{Re}(\beta_j) r}+\delta_{\text{Im}(\beta_j) r})\delta_{rs}\delta(t-t')$ to probe the stability of the attractor for $1000\cdot \Omega_2/(2\pi) \approx 1150$ oscillation periods of $\Omega_2$. The step size $\delta t=0.0001$ throughout every simulation is chosen to be such that we have approximately 11500 sample points per oscillation period of $\Omega_2$ in order to numerically converge.
We conducted a larger parameter scan of the modulation depth $d$ and the modulation frequency $\Omega_{\text{mod}}$ to understand the effects of the modulated drive. The numerical parameters are identical to the examples shown in Fig.~1 of the main text, apart from the modulation depth which is scanned from 0 to 0.1 in steps of 0.005, and the modulation frequency which is swept from 0 to 3 in steps of 0.05. The results are summarized in Fig.~\ref{fig:PhaseD}. The classification of a simulation into a thermal state (black square), self-sustained oscillation of one mode (blue square) or multimode oscillation (red square) is based on the Fourier spectra of the mechanical displacements $\text{Re}(\beta_j)$ which were computed for $980\cdot \Omega_2/(2\pi) \approx 1100$ oscillation periods of $\Omega_2$. A simulation is classified as a thermal state if that all amplitudes in the range $\omega \in [4,10]$ are below the threshold value $10^{10}$ which can be seen in the inset of Fig 1.(b). Self-oscillation and multimode oscillation are related to a rise of the amplitude for either one or both mechanical displacements around their natural frequency above the threshold value $10^{11}$. In addition, to classify a simulation as multimode oscillation, the maximal amplitude in the spectrum of the mechanical displacement of $\beta_2$ needs to be in the interval $[6,8]$ and a fit of the logarithm of a Lorentzian function $f(\omega)=\ln(\tilde{S}_0+a[(\omega-\tilde{\Omega}_2)^2+(\Gamma'_2/2)]^{-1})$ to the logarithm of the Fourier spectrum in that range results in a narrowed linewidth of $\Gamma'_2<\Gamma_2/2$ while keeping the frequency $\tilde{\Omega}_2=6.8$ fixed. We see that increasing the modulation depth takes the system below laser threshold for low modulation frequencies. This can be understood since the total power that exits the Mach--Zehnder modulator is constant independent of the modulation depth and some of the power is taken from the main tone at $\omega_L$ to generate the two sidebands at $\omega_L\pm\Omega_{\text{mod}}$, which are not optimally detuned to generate lasing. If the modulation frequency is swept towards the difference frequency we see that the additional stimulated emission process first helps one mode to overcome the oscillation threshold and at the difference frequency and beyond the second mode also surpasses the threshold. The time requirements of the numerical algorithm limit the amount of simulations considered per data point leading to statistical fluctuations in the phase diagram. The white stars denote the modulation parameters for the shown simulations in Fig.~1 of the main text.

\section*{V. The role of the optical spring effect in frequency stability}
One effect that could play a role in the observed frequency stabilization in the MMO state is related to fluctuations transduced through the optical spring effect. To understand this in more detail, consider the central frequency corrected by the optical spring shift in Eq. (10) in the main text
\begin{align}
\Omega'_j(\Omega_j)=\Omega_j\sqrt{1-\frac{\bar{\Delta} g_j^2|\bar{\alpha}_0|^2[\frac{\kappa^2}{4}-\Omega_j^2+\bar{\Delta}^2]}{\Omega_j[\frac{\kappa^2}{4}+(\Omega_j-\bar{\Delta})^2][\frac{\kappa^2}{4}+(\Omega_j+\bar{\Delta})^2]}}.
\end{align}
Observing that $|\Omega'_j-\Omega_j|\ll\Omega_j$, we apply the Taylor expansion $\sqrt{1-ax^2} \approx 1-\frac{ax^2}{2}$ leading to the approximated central frequency
\begin{align}
\Omega'_j(\Omega_j) \approx \Omega_j-\frac{\bar{\Delta} g_j^2|\bar{\alpha}_0|^2[\frac{\kappa^2}{4}-\Omega_j^2+\bar{\Delta}^2]}{2[\frac{\kappa^2}{4}+(\Omega_j-\bar{\Delta})^2][\frac{\kappa^2}{4}+(\Omega_j+\bar{\Delta})^2]}.
\end{align}
A change of laser intensity $|\bar{\alpha}_0|^2$ by an amount $\delta I$ leads to a change of the mechanical frequency 
\begin{align}
\delta \Omega'_j = -\frac{\bar{\Delta} g_j^2[\frac{\kappa^2}{4}-\Omega_j^2+\bar{\Delta}^2]}{2[\frac{\kappa^2}{4}+(\Omega_j-\bar{\Delta})^2][\frac{\kappa^2}{4}+(\Omega_j+\bar{\Delta})^2]} \delta I.
\end{align}
As such, laser intensity fluctuations, either quantum fluctuations or classical fluctuations related to the laser or the tapered optical fiber, could affect the stability of the mechanical resonator through the induced optical spring shift. Moreover, Uncertainty in other quantities such as the detuning $\bar{\Delta}$ contributes with additional error. The thermo-optic effect, which leads to laser-induced cavity red-shift in the silicon nanocavities we employ, stabilizes fluctuations in intensity and detuning only to a finite degree. The magnitude of the resulting spring shift depends on the mean intracavity photon number. A reduction of the intracavity photon number as suggested by the theoretical finding that the lasing threshold power is lowered by the modulation as seen in the phase diagram, and the observed reduction of the peaks in the experimental spectra in the MMO state, make it possible that the observed stabilization is at least in part related to this mechanism.

\section*{VI. Experimental Setup}
\begin{figure}[h]
\begin{center}
\includegraphics[width=0.45\textwidth]{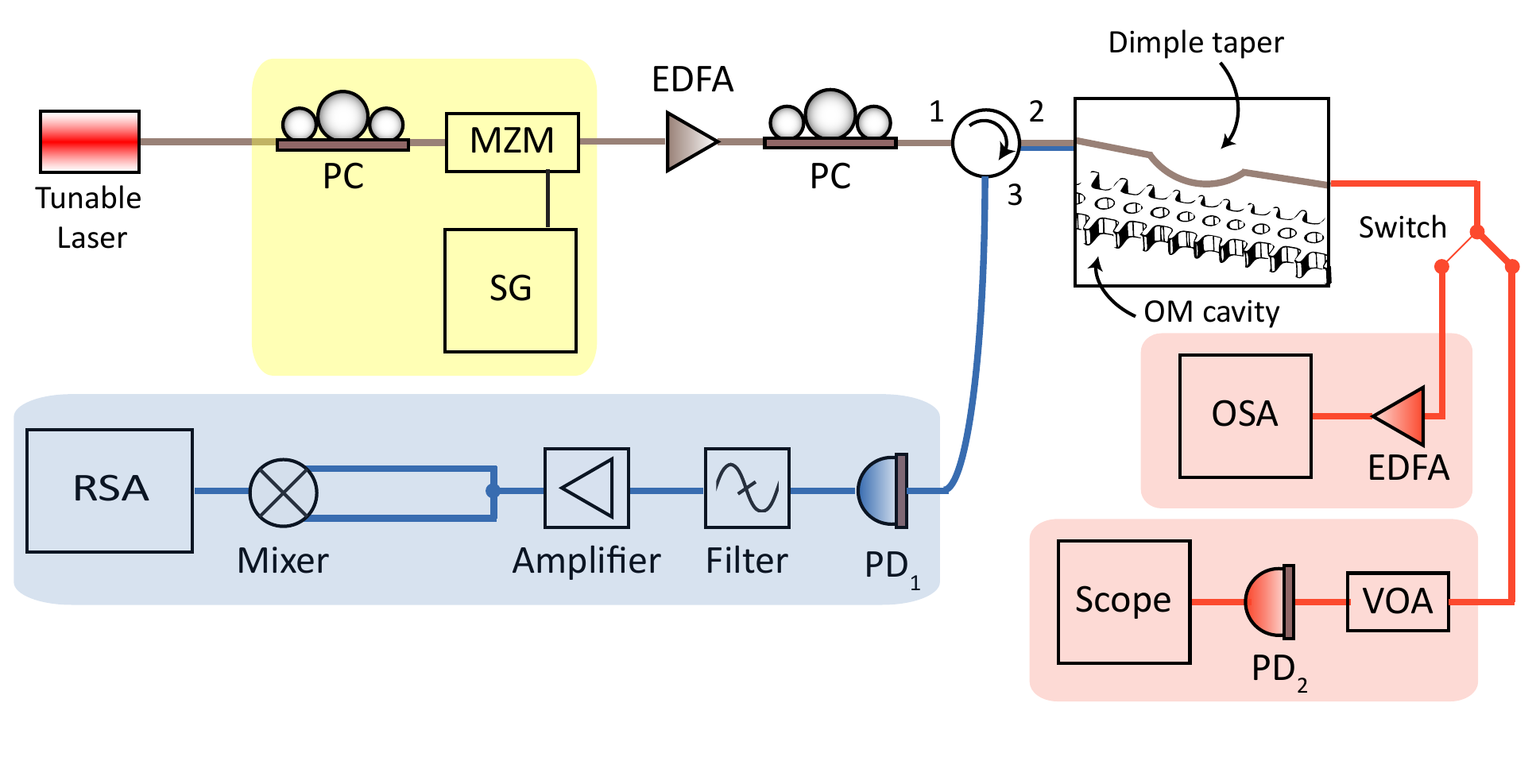}%
\end{center}
\caption{Schematic of the experimental setup used in the multimode experiment lasing.}
\label{fig:setup}
\end{figure}

The measurements were performed with the experimental setup illustrated in Fig.~\ref{fig:setup}. Two similar setups were installed (independently) at the AMOLF institute in the Netherlands and the Nanophotonics Technology Center (NTC) in Spain, where different compatible and reproducible sets of measurements were obtained. A tunable fiber-coupled external cavity diode laser (New Focus TLB-6728) generates a continuous-wave optical signal that passes through an intensity modulator fed by a signal generator (SG) with a tuneable modulation frequency $\omega_{RF}$. In both setups the used intensity modulator was a Covega Mach MZM and the SG was an EXG Analog Signal Generator N5173B at AMOLF and an Agilent E4438C ESG Vector Signal Generator at NTC. The modulated laser signal is passed through an erbium doped fiber amplifier (EDFA), a polarization controller (PC), and an optical circulator before it is sent into a dimpled tapered fiber. Another employed version of the system consists in coupling the light into and out of the cavity with a fiber taper loop. When the dimpled fiber is close enough to the OM cavity under study, light is coupled evanescently from the fiber to the cavity so that we can characterize both the transmission and reflection spectra, shaded/highlighted in red and blue, respectively.

The transmitted signal is first sent to a variable optical attenuator (VOA), where it is photodetected ($PD_{1}$) to monitor the optical resonance with an oscilloscope (`Scope'). The reflected signal is photodetected after the circulator via a 12 GHz Photodetector (New Focus 1544-B DC-Coupled NIR Fiber-Optic Receiver) ($\text{PD}_{2}$) and then filtered with a band pass filter (Mini-Circuits VBFZ-4000-S+). This filter is used to suppress the modulated frequency $\omega_{RF}$ in the detected electrical signal which would overlap with the difference tone. Once the low frequency range has been filtered, the signal is electrically amplified (Miteq MPN4-02001800-23P) and divided (Mini-Circuits 15542 ZFRSC-42) in order to feed the two input ports of an electrical mixer (MACOM M63C), where the difference and sum tones of the two lasing mechanical modes P1 and P2 are created. Finally, the resulting signal is analyzed with a radiofrequency spectrum analyzer (RSA). The RSA used at AMOLF was a MXA Signal Analyzer N9020A and at NTC a Aniritsu MS2850A Signal Analyzer. All the phase noise measurements presented in this work were obtained with the latter one. Real-time spectrograms of the spectrum versus time were acquired on both. 

\begin{figure}[h]
\begin{center}
\includegraphics[width=0.45\textwidth]{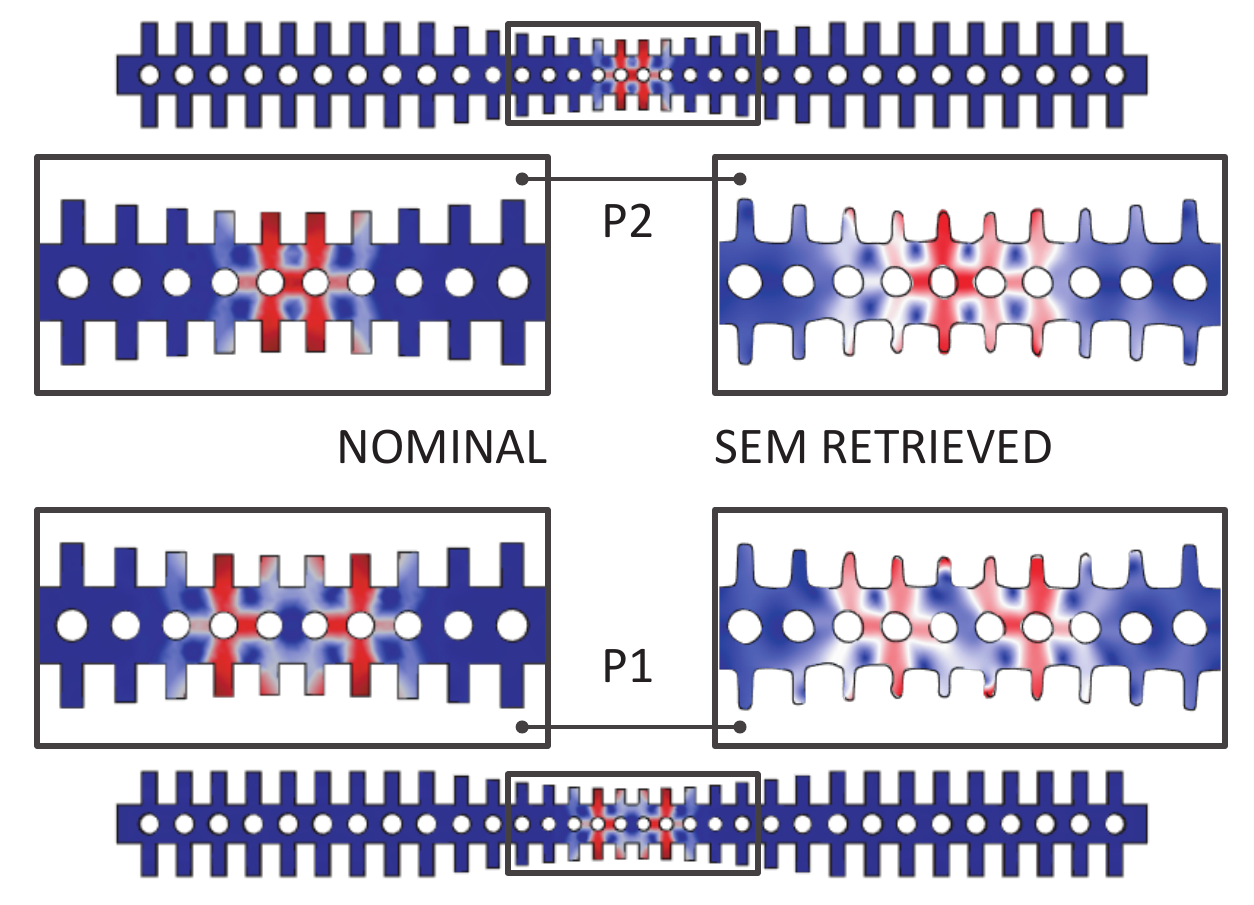}%
\end{center}
\caption{Comparison of the mechanical mode profile of the nominal structures and the final fabricated profile structure for the two mechanical modes involved in the experiments.}
\label{fig:modes}
\end{figure}

To reach the MMO regime the experiment was performed as follows: First, the difference frequency between the two involved mechanical modes was characterized at different laser wavelengths. Next, this difference frequency was set as the modulation frequency of the laser driving the cavity, and it was kept constant in the rest of the experiment. Finally, a sweep of the laser wavelength on the blue-detuned side of the resonance was performed, reaching in this way the self-oscillation regime in the MMO state. 

The difference tone $\Omega_2-\Omega_1$ is also measured with this experimental setup to confirm the mode-locking between P1 and P2. This is achieved by sending the detected signal through the microwave mixer. Using a control measurement, we verified that the filter fully suppresses the (low-frequency) optical modulation frequency recorded by the photodetector, such that the measured difference-frequency signal was only related to the mixing of the two GHz-frequency signals of the oscillators P1 and P2.

\section*{VII. Mechanical mode characterization}
Because of fabrication imperfections, the final structures differ from the nominal ones. In order to ensure that the measured mechanical modes were still located in the middle region of the cavity, a calculation of the mechanical modes was performed using a profile retrieved from a Scanning Microscopy Image (SEM). In Fig.~\ref{fig:modes} both the theoretical mechanical mode profile of the nominal structure and the real profile fabricated structure are compared. Here, we can see similarity between the two profiles of the mechanical modes in both cases.

\begin{figure}[h]
\begin{center}
\includegraphics[width=0.5\textwidth]{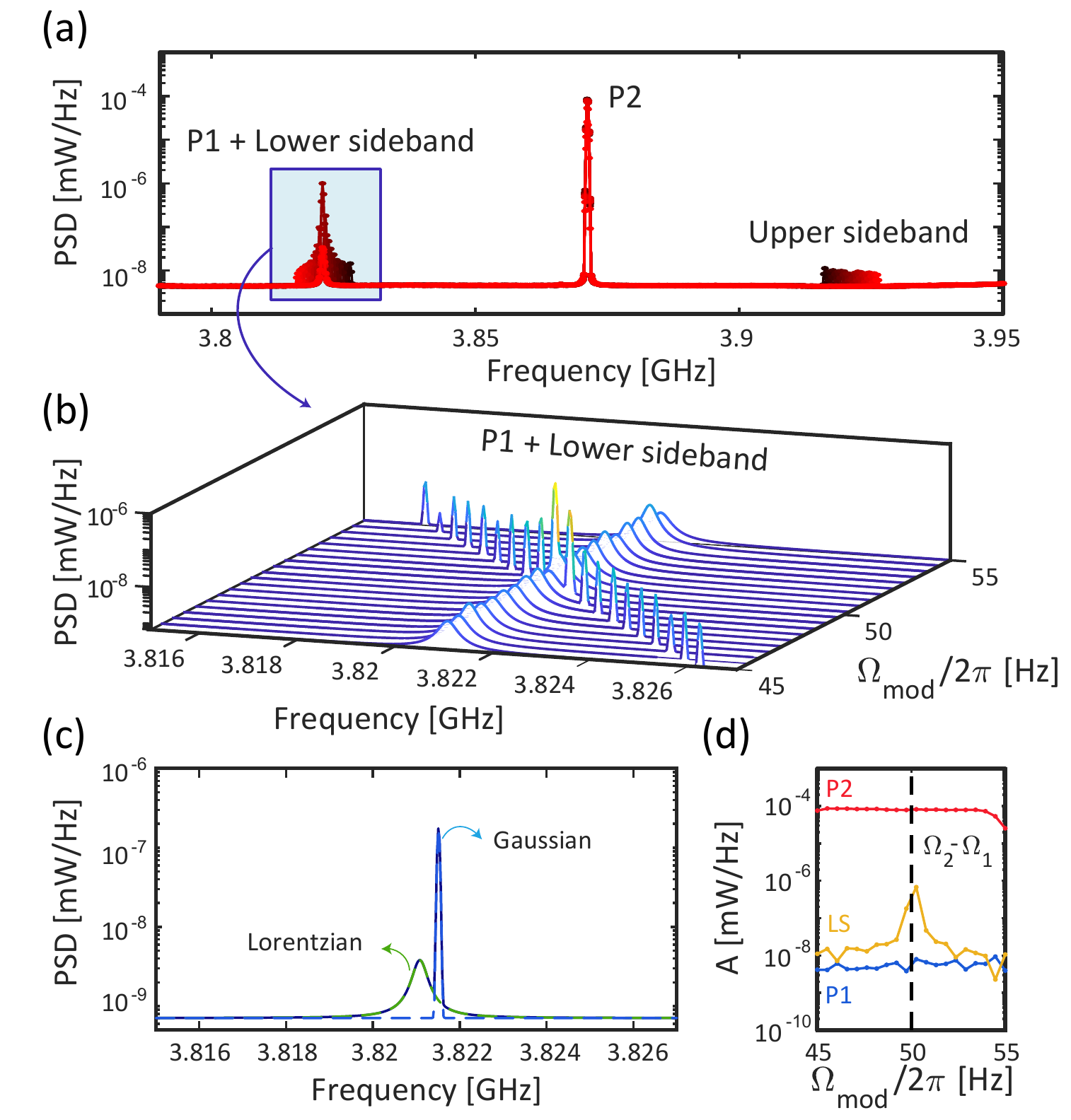}%
\end{center}
\caption{Modulation frequency scan. (a) Overview of the upconversion of the modulation tone around $\Omega_{2}-\Omega_{1}$. (b) Close view of the scan of the modulation tone around P1. (c) Example of the different fit contributions in the overlap between P2 and the lower sideband. (d) Evolution of the amplitudes of P1, P2 and the lower sideband as a function of the modulation frequency tone.}
\label{fig:modulationfreq}
\end{figure}

\section*{VIII. Modulation frequency scan}
Once P1 or P2 is in the lasing state, an upconversion of the drive modulation tone occurs. This means that the drive tone gives rise to a lower and an upper sideband of the lasing tone in the optical modulation spectrum. This can be seen in Fig.~\ref{fig:modulationfreq}(a), which shows many spectra (colored from black to red) that are taken while varying the modulation frequency around the difference frequency of P1 and P2. A close view of this scan is shown in Fig.~\ref{fig:modulationfreq}(b). Here, we can see that for modulation frequencies around $\Omega_{2}-\Omega_{1}$, the lower sideband (LS) overlaps with the mechanical mode P1. In order to observe whether the mechanical modes P2 or P1 experience a change, we analyzed the amplitude of those modes and the lower sideband. It has to be noted that in the case of the overlap between P1 and the lower sideband, we have performed a fit of a Lorentzian in the case of the thermally driven mechanical mode P1 and a fit of a Gaussian in the case of the lower sideband driven by modulation tone as can be seen in Fig.~\ref{fig:modulationfreq}(c). We find that the linewidth of the Lorentzian (P1) is never narrowed while the modulation sideband of P2 is scanned in and out of resonance with mode P1. This shows that the MMO state cannot be reached from the SSO state. The resulting evolution of both P1, P2 and LS amplitudes are presented in Fig.~\ref{fig:modulationfreq}(d).

\begin{figure}[h]
\begin{center}
\includegraphics[width=0.5\textwidth]{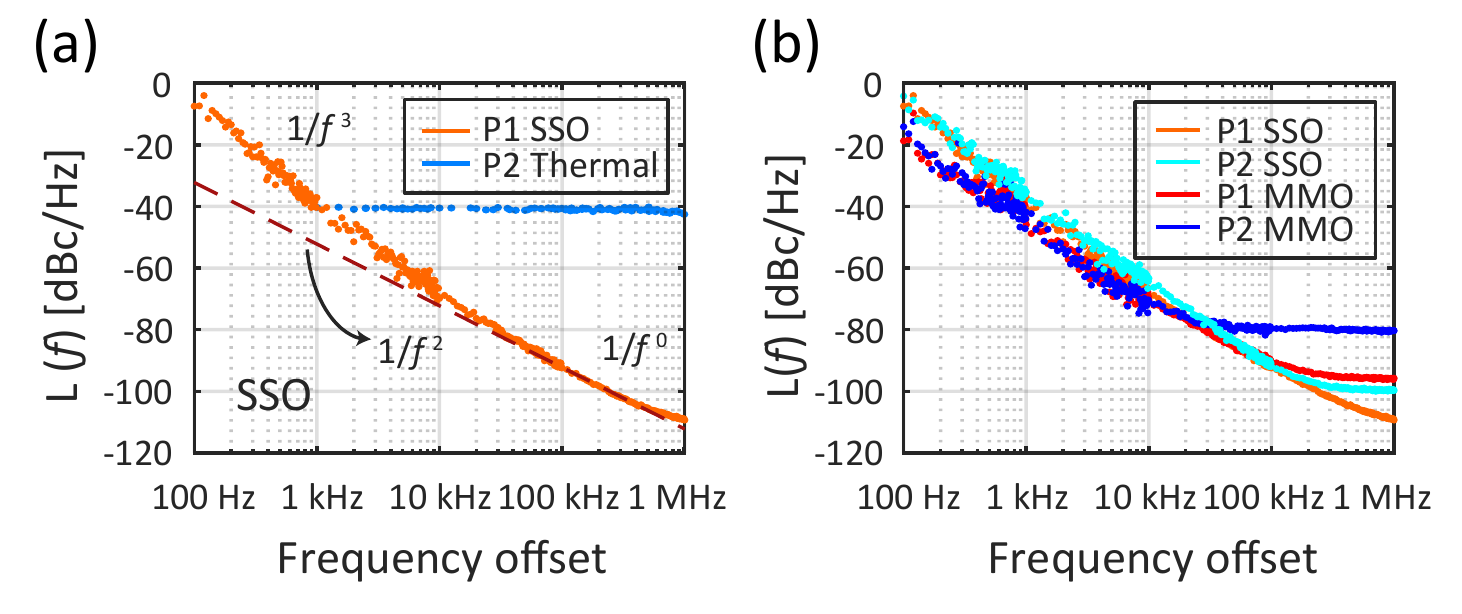}%
\end{center}
\caption{(a) Phase noise $L(f)$ from a thermally driven mechanical mode and a SSO mode. (b) Comparison of the two mechanical modes P1 and P2 in the SSO and MMO states.}
\label{fig:phasenoiseall}
\end{figure}

\section*{IX. Phase noise measurements}
The phase noise measurements from 100 Hz to 1 MHz frequency offset were performed with an Aniritsu MS2850A Signal Analyzer, as mentioned above. A measurement of the phase noise of a mechanical mode thermally driven and a SSO mode was performed in order to be able to distinguish both situations. These measurements are presented in Fig.~\ref{fig:phasenoiseall}(a) for the case where P1 is self-oscillating. Here, we can see an appreciable difference between these two cases as the thermally driven mode had a very low amplitude, thus resulting in a huge contribution of frequency noise sources related to the white phase noise. 

On the other hand, Fig.~\ref{fig:phasenoiseall}(b) shows a comparison of the phase noise for all the cases of the mechanical modes P1 and P2. P1 SSO and P2 SSO were taken when the corresponding mechanical modes were in the self-oscillation state and P1 MMO and P2 MMO were taken when both of them were simultaneously oscillating. As discussed in the main text, the most interesting feature arises in the low frequency offset regime, where we can see a difference in the phase noise from SSO and MMO. For all frequencies below $\sim$10 kHz (where the white noise background is insignificant), the phase noise improves when the two mechanical modes are in the multimode regime. For these frequencies, the phase noise has a largest contribution related to flicker frequency noise, judging from the slope of this curve.

\section*{X. Allan deviation calculation}
The calculation of the Allan deviation $\sigma_{y}(\tau)$ was performed in two different ways. For small averaging times $\tau$, the parameter was derived from the phase noise data as \cite{Luan2014}:
\begin{equation}
\sigma_{y}(\tau)=\sqrt{\int_{0}^{\infty}\frac{4f^{2}L(f)}{f_{c}^{2}} \frac{\sin^{4}(\pi\tau f)}{(\pi\tau f)^{2}}\text{d}f},
\end{equation}
where $L(f)$ is the measured phase noise and $f_{c}$ the carrier frequency of the oscillator under study. 

\begin{figure}[h]
\begin{center}
\includegraphics[width=0.5\textwidth]{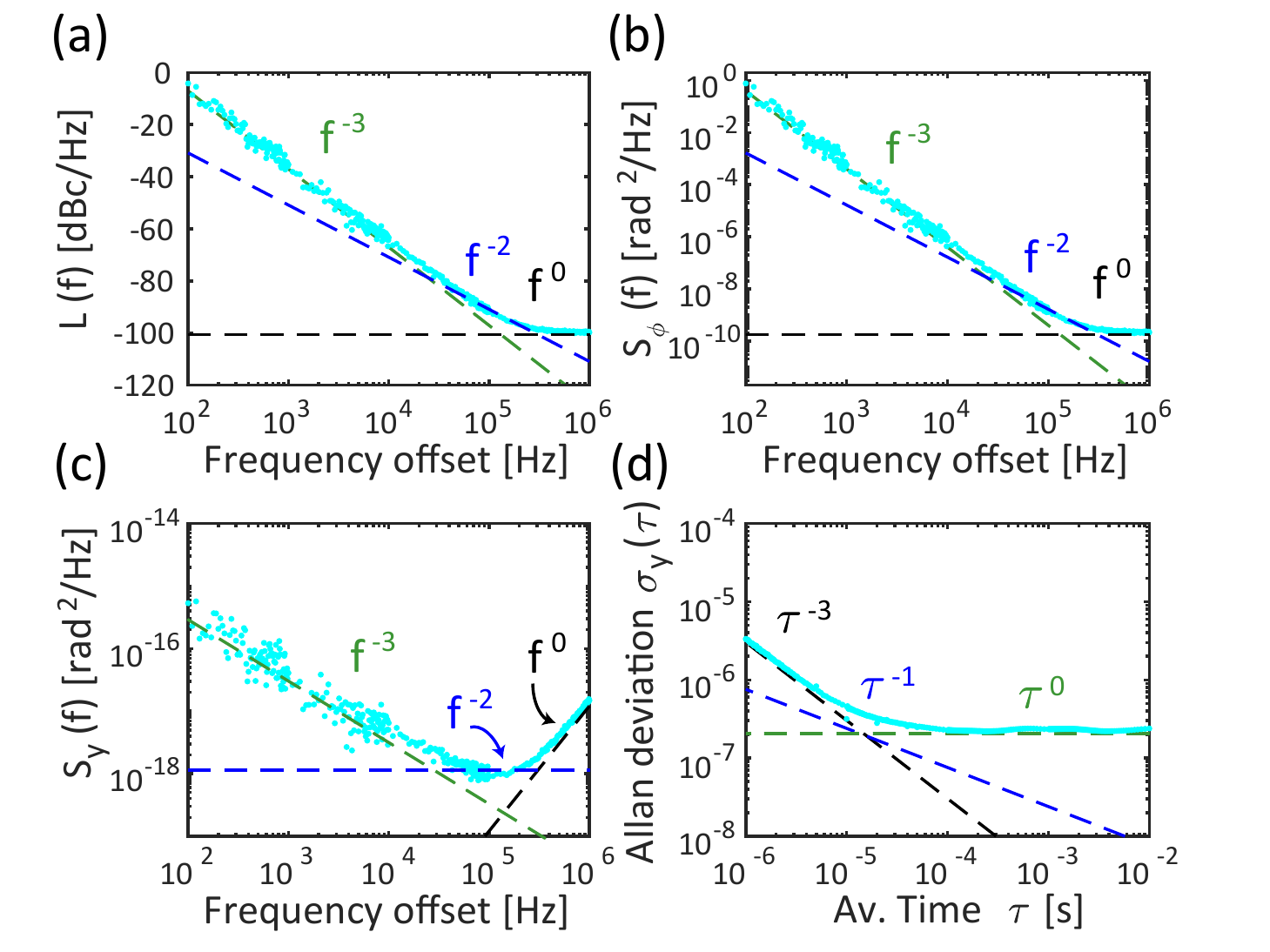}%
\end{center}
\caption{P2 SSO Noise type contributions. (a) Phase noise $L(f)$. (b) Phase noise $S_{\varphi}(f)$. (c) Frequency noise $S_{y}(f)$. (d) Allan deviation $\sigma(\tau)$. The phase noise fit contributions described by Leeson's model for the frequency and time domain are indicated with dashed lines.}
\label{fig:phasenoisecontri}
\end{figure}

Figure \ref{fig:phasenoisecontri}(a) shows the common SSB phase noise $L(f)$. However, the most common quantity to describe the oscillator phase noise is $S_{\varphi}(f)$ which can easly be derived as $S_{\varphi}(f)=2L(f)$. In order to derive $\sigma(\tau)$ an auxiliary parameter $S_{y}(f)$ is also needed, which is related to the physical parameter $S_{\varphi}(f)$ as
\begin{equation}
    S_{y}(f)=\dfrac{f^{2}}{f_{c}^{2}}S_{\varphi}(f).
\end{equation}

The phase noise shows the typical dependencies, with $1/f^{3}$ (flicker frequency noise, lower part of the spectrum Fig.~\ref{fig:phasenoisecontri}(a,b)), $1/f^{2}$ (white frequency noise, upper part of the spectrum in Fig.~\ref{fig:phasenoisecontri}(a,b)) and white noise $1/f^{0}$, which are in good agreement with the general phase noise described by the Leeson's model \cite{Rubiola, Mercade2020}. 

Besides the phase noise as a measure about the stability of a signal, we can also study the root mean square (RMS) jitter ($J_{RMS}$). It can be obtained by integrating the phase noise power data as \cite{Luan2014}:

\begin{equation}
J_{RMS}=\frac{1}{2\pi\nu_{0}}\sqrt{2\int_{f_{1}}^{f_{2}}10^{L(f)/10}df}
\label{eq:jitter}
\end{equation}

\noindent where $f_{1}$ and $f_{2}$ are the start and stop frequency, respectively. 

On the other hand, for the long-term stability another approach was followed, in order to characterize the frequency drift of our oscillators. Besides the approximation of the calculation of the Allan deviation from the phase noise, this stability measurement can also be described as \cite{Yu2016}
\begin{align}
\sigma(\tau)=
\sqrt{\dfrac{1}{2M}\sum_{i=0}^{M-1}(y_{i+1}-y_{i})^{2}}
\end{align}
where $M=T/\tau-1$ and $y_{i}=(\langle f_{1}(t_{0}+i\tau)\rangle_{\tau}-f_{c})/f_{c}$. Here, $\langle f_{1}\rangle_{\tau}$ is the average frequency of the system over the interval $\tau$.

In our experiment, the measurement of the evolution of the frequency as a function of time was performed with a real time electrical spectrum analyzer over a total measurement time of 1 s and a timing step around 1 ms.

\end{document}